\documentclass{jfm}

\usepackage{subcaption}
\usepackage{amsmath}
\usepackage{MnSymbol}
\usepackage{color}

\shorttitle{ERPP method for wall-bounded flows}
\shortauthor{F. Battista, J.-P. Mollicone, P. Gualtieri, R. Messina and C.M. Casciola}

\title[ERPP method for  wall-bounded flows]{Exact regularized point particle (ERPP) 
method for particle-laden wall-bounded flows in the two-way coupling regime}

\author{F. Battista$^1$\corresp{\email{francesco.battista@uniroma1.it}}, J.-P. Mollicone$^2$, P. Gualtieri$^3$, R. Messina$^3$ \and C.M. Casciola$^3$}

\affiliation{
$^1$ENEA, Italian Agency for New Technologies, Energy and Sustainable Economic Development, via Anguillarese 301, 00123 Rome, Italy\\
$^2$Department of Civil and Environmental Engineering, Imperial College London, United Kingdom\\
$^3$Department of Mechanical and Aerospace Engineering \\ Sapienza University of Rome \\ via Eudossiana 18, 00184 Rome, Italy}

\newcommand {\bean}  {\begin{eqnarray*}}
\newcommand {\eean}  {\end{eqnarray*}}

\newcommand {\ttau}   {\tilde \tau}

\def\Rset {{\rm I \kern-.2em R}} 

\newcommand {\bce}  {\begin{center}}
\newcommand {\ece}  {\end{center}}
\newcommand {\be}   {\begin{equation}}
\newcommand {\ba}   {\begin{array}}
\newcommand {\bea}  {\begin{eqnarray}}
\newcommand {\bfi}  {\begin{figure}}
\newcommand {\ee}   {\end{equation}} 
\newcommand {\ea}   {\end{array}}
\newcommand {\eea}  {\end{eqnarray}}
\newcommand {\efi}  {\end{figure}}

\newcommand {\vpi}  {\mbox{\boldmath $\pi$}}

\newcommand {\vxi}    {\mbox{\boldmath $\xi$}}
\newcommand {\vzeta}  {\mbox{\boldmath $\zeta$}}

\newcommand {\vomega} {\mbox{\boldmath $\omega$}}

\newcommand {\vD}     {{\bf D}}
\newcommand {\ve}     {{\bf e}}

\newcommand {\vf}     {{\bf f}}
\newcommand {\vF}     {{\bf F}}

\newcommand {\vI}     {{\bf I}}

\newcommand {\vn}     {{\bf n}}

\newcommand {\vu}     {{\bf u}}

\newcommand {\vv}     {{\bf v}}
\newcommand {\vV}     {{\bf V}}
\newcommand {\vx}     {{\bf x}}

\newcommand {\vw}     {{\bf w}}

\def\0{{_0}}
\def\Rset {{\rm I \kern-.2em R}} 
\def\Eset {{\rm I \kern-.2em E}} 
\def\mathbbH {{\rm I \kern-.2em H}} 
\def\mathbbC {{\rm I \kern-.6em C}} 

\begin{document}
\maketitle
\begin{abstract}

The Exact Regularized Point Particle (ERPP) method is extended to treat the inter-phase 
momentum coupling between particles and fluid in the presence of walls by accounting for 
the vorticity generation due to the particles close to solid boundaries. The ERPP method 
overcomes the limitations of other methods by allowing the simulation of an extensive parameter 
space (Stokes number, mass loading, particle-to-fluid density ratio and Reynolds number) and of 
particle spatial distributions that are uneven (few particles per computational cell). The enhanced 
ERPP method is explained in detail and validated by considering the global impulse balance. 
In conditions when particles are located close to the wall, a common scenario in wall-bounded turbulent 
flows, the main contribution to the total impulse arises from the particle-induced vorticity at the solid 
boundary. The method is applied to direct numerical simulations of particle-laden turbulent 
pipe flow in the two-way coupling regime to address the turbulence modulation. The effects of the 
mass loading, the Stokes number and the particle-to-fluid density ratio are investigated. 
The drag is either unaltered or increased by the particles with respect to the uncoupled case. 
No drag reduction is found in the parameter space considered. The momentum stress budget, 
which includes an extra stress contribution by the particles, provides the rationale behind the 
drag behaviour. The extra stress produces a momentum flux towards the wall that strongly modifies 
the viscous stress, the culprit of drag at solid boundaries.

\end{abstract}
\section{Introduction} \label{sec:introduction}
Particle laden turbulent flows are ubiquitous and  challenging due to the multi-scale 
physics involved, see \cite{vanderhoef_rev}. 
Turbulence has an important role in the motion of particles. 
The transport, entrainment and redeposition, 
\cite{soldati_rev}, of solid particles, such as coal dust, is crucial
in determining the overall efficiency of energy plants, \cite{buhre2005oxy}.
In automotive applications, the spray formation, \cite{Marmottant_2004}, and the 
ensuing fuel jets, \cite{jenny}, impact the overall efficiency of combustion.
Physical phenomena such as inter-particle collisions,
\cite{post,wang_rev}, and turbulent modification, \cite{balachandar_rev}, play  
an important role. In many applications, the multi-scale nature of the
phenomena involved calls for modelling both the fluid turbulence and the
particle motion, see \cite{meyer2012modelling,peirano2006mean} for 
modelling strategies in Reynolds Averaged Navier Stokes (RANS) equations and 
\cite{marchioli2017large,innocenti2016lagrangian} 
in Large Eddy Simulation (LES). 

In the context of particle laden turbulent flow many studies have been conducted in
the one-way coupling regime, see \cite{elghobashi1994predicting} for a discussion 
of the different regimes of coupling between fluid and particles, both from 
experimental point of view see e.g. \cite{kostinski2001scale,lau2016effect} and 
\cite{eidelman2009mixing}, and from numerical point of view, 
see e.g.~\cite{toschi2009lagrangian,bec2007heavy,sardina2012self,
picano2011dynamics,marchioli2002mechanisms,goto2006self} 
and \cite{battista2011intermittent}.
On the other hand, the two-way coupling regime, where the fluid/particle 
momentum exchange is significant, is still being thoroughly investigated and some 
open questions need to be addressed. The first one is related to the 
numerical technique employed to model a reliable fluid/particle interaction. 
The second is related to the physics and deals with the particle dynamics, 
their spatial distribution and most importantly, with the turbulence modulation.

In the literature, different techniques are used to model the fluid/particle 
interaction. The approaches mostly depend on the typical size of the 
particle, e.g. the diameter $d_p$, that has to be compared with the 
characteristic length scales of the turbulent flow, 
{see e.g. the recent review paper by \cite{elghobashi2019direct}.}
The conceptually simplest approach is to resolve the particle 
boundary on the computational grid and to enforce the no-slip conditions
on each particle boundary (particle-resolved simulation).
This technique has recently become viable due to the 
ever-increasing computing resources. Among many, the 
immersed boundary technique,~\cite{uhlmann2005immersed,breugem2012second}, 
is employed to simulate suspensions under different conditions, for example in 
sedimentation problems, \cite{fornari2016rheology,fornari2016sedimentation}, or 
dense suspensions, \cite{costa2018effects}. Given the tremendous computational 
cost, such simulations can currently only tackle problems in simplified geometries 
where the scale of the particle is roughly 10 times larger than the 
dissipative scale of the turbulent flow and the choice of density ratio is limited.

In many applications, the typical particle diameter is comparable to, or even 
smaller, than the dissipative scale and the density ratio is relatively large. 
In these conditions, it is unaffordable to carry out particle-resolved simulations. 
Being small, the particles are modelled as material points 
which behave as concentrated momentum sources/sinks for the fluid via the 
hydrodynamic drag that the (small) particle experiences along its trajectory. 

In wall-bounded turbulent flows, the particles segregate towards the wall, 
\citep{caporaloni1975transfer,young1997theory}. The phenomena is known 
as turbophoresis and is relevant since the particles close to the wall 
affect the turbulence in the buffer region where the production of turbulent 
kinetic energy occurs together with the generation of the vortical structures,
see~\cite{bijlard2010direct,dritselis2008numerical,dritselis2011numerical} 
and \cite{richter2014modification} for the analysis of the topological 
modification of the structures in the buffer region.

An important issue in wall-bounded turbulent flows is whether the 
disperse phase feedback produces an overall increase or decrease 
of the friction and whether the turbulent fluctuations are augmented or reduced 
by the particles. \cite{zhao2010turbulence,zhao2013interphasial} 
found reduction of the friction, i.e. the flow rate in presence of particles 
is augmented with respect to the flow rate in absence of particles for
the same pressure gradient that drives the turbulent channel flow.
\cite{picciotto2006turbulence} and \cite{li2016modulation,li2016direct} 
considered the turbulence modulation in a boundary layer,
showing an increase in the skin friction coefficient at the wall.
\cite{li2001numerical} report that the overall drag might increase or decrease 
depending on the mass loading of the suspension and the particle Stokes number. 
\cite{lee2015modification} found an overall increase of the turbulent fluctuations 
for relatively small particles, and a decrease of the turbulent 
fluctuations for large particles. In contrast, \cite{pan1997numerical} found an 
overall increase in drag in a channel flow.

All the simulations mentioned above exploit the classical Particle In Cell (PIC) approach 
introduced by~\cite{crowe}, except for \cite{pan1997numerical} who employed an alternative 
inter-phase momentum coupling based on the solution of a truncated steady Stokes 
flow for the disturbance flow produced by the particles. Simulations using PIC 
have been performed also in the pipe flow, see e.g. the 
DNS by \cite{rani2004numerical} and \cite{vreman2007turbulence}, and Large-Eddy 
Simulation (LES) by~\cite{yamamoto2010numerical}. Recently, the effect of the 
wall roughness has been discussed by \cite{vreman2015turbulence} 
and \cite{de2016turbulence}.

From the experimental point of view, the motion, deposition, 
entrainment, spatial distribution and velocity profiles of the particles in 
a turbulent boundary layer have been addressed 
in \cite{kaftori1995particleI,kaftori1995particleII} and \cite{kaftori1998effect}.
The fluid velocity profiles show larger gradients close to the wall 
(drag increasing) and the turbulent velocity fluctuations are increased in 
the near wall region. These modifications are associated with an increase in 
wall shear stress. Similar results are found in the experiments 
by~\cite{wu2006experimental,li2012experimental} 
and~\cite{righetti2004particle}. No substantial modification of the 
mean velocity profile has been reported by \cite{kulick1994particle} where
only turbulent fluctuations are depleted across the channel.
In the geometry of the pipe, \cite{tsuji1984ldv} show an increase in wall 
shear stress as well as \cite{hadinoto2005reynolds}. See also other experiments 
by \cite{caraman2003effect,boree2005dilute} and \cite{ljus2002turbulence}.

In the numerical simulations discussed, the inter-phase momentum coupling is 
mainly achieved using the PIC approach. Even though the approach is rather simple, 
it suffers from several drawbacks. Firstly, the backreaction field, that can be constructed 
given the configuration of the suspension, is grid dependent~\citep{jfm_2way}. Secondly, 
the solution depends on how many particles per computational cell are available, 
see~\cite{jfm_2way,boivin1998direct} and the general discussion by~\cite{balachandar_rev}. 
The unphysical constraint on the number of particles per cell poses several limitations 
on the range of the dimensionless parameters (Stokes number, mass loading, particle-to-fluid 
density ratio and Reynolds number) that can be explored in the simulations, see the conclusion 
of~\cite{gualtieri2015exact}.
A further issue concerns the model required to compute the hydrodynamic force on each particle. 
In the simple model of the Stokes drag, the fluid velocity at the particle position must be correctly 
interpreted as the background fluid velocity, i.e. as the fluid velocity in absence of the disturbance 
flow produced by the specific particle under consideration. Unfortunately, in two-way coupled 
simulations, the unperturbed flow is unavailable unless specific techniques are exploited to 
remove the particle self-disturbance, see e.g.~\cite{horwitz2016accurate,horwitz2017correction,capecelatro2013euler}
and \cite{ireland2017improving,akiki2017pairwise} where several methods are proposed to circumvent this problem.
{These considerations pose challenging issues from the theoretical point of view and 
call for more accurate modelling of the particle/fluid interaction. The Exact Regularized Point Particle (ERPP) 
method has been proven to correctly evaluate the particle hydrodynamic force since the particle self-disturbance flow is 
known in a closed form. Moreover, the approach provides convergent turbulent statistics at the smallest scales
of the flow, see~\cite{gualtieri2017turbulence,battista2018application} }.

The aim of the present manuscript is to generalise the ERPP method, originally derived for free space flows, 
for the simulation of particle-laden wall-bounded turbulent flows and to provide a parametric study of turbulence
modulation in conditions inaccessible to the classical PIC method. 
{A treatment of the particle phase which is not sufficiently accurate would produce an incorrect force field 
on the fluid, mostly in a narrow layer close to the bounding walls thus altering the delicate balance of 
momentum in the wall layer and may lead to unphysical macroscopic effects. 
Once particles form clusters and segregate near the wall, the force they exert on the fluid will depend on 
the cluster geometry and, since clusters are generated by the small turbulent scales, a non-convergent algorithm 
will poorly reproduce the overall physics, i.e. the modification of wall friction due to the particles.
The extended ERPP method overcomes these issues and provides a systematic approach to accurately predict the
dynamics of wall-bounded particle-laden turbulent flows free of numerical artefacts. 
After the basic dynamics is captured, more sophisticated observables
can be addressed and trusted given the convergence properties of the approach, in order to address
higher order statistics in shear dominated flows, \cite{jacob2008scaling}, or the scale-by-scale dynamics,
\cite{mollicone2018turbulence}.}

Technological applications generally involve turbulent flows in a complex geometry. 
In view of providing a sufficiently general approach, it is important to include the effects 
of wall curvature when studying the inter-phase momentum exchange between particles and fluid 
close to solid boundaries. For this reason, the simplest flow configuration that can be addressed 
is the turbulent flow inside a circular pipe, which we simulate using direct numerical simulation.
Most of the turbulent fluctuations are generated in the near wall region, 
see e.g.~ \cite{marusic2010wall,marusic2013logarithmic,mathis2009large,hwang2010self},
requiring an accurate modelling of the inter-phase momentum exchange close
to solid walls. 

The ERPP method enables a free choice of the control parameters, that is the Stokes number, 
mass loading, particle-to-fluid density ratio and Reynolds number, allowing to explore a region of 
the parameter space which is not possible for other approaches, such as the PIC method and 
resolved particle method. For example, when the particle-to-fluid density ratio is order $20 - 200$, 
the number of particles turns out to be small for a given grid resolution imposed by the Reynolds number. 
These values roughly correspond to cases involving medicinal particulate commonly used for inhalable drug 
delivery systems, carbon dust transport, food industry powders resulting from the processing of cereals and 
sawdust resulting from wood manufacturing. Another advantage arises when the carrier phase is 
relatively dense, such as water, resulting in relatively low density ratios considering common materials. 
The ERPP method also allows the simulation of  flows where the particle spatial distribution is uneven 
and few particles per cell are found in some regions of the flow. This may occur, for example, in water 
steam flows where the condensation of small droplets, imposed by external thermodynamical conditions, 
dictates the number of particles in a specific flow region. 
 
The paper is organised as follows: section~\ref{sec:methodology} provides the theoretical background of 
the methodology for wall-bounded flows. Section \ref{sec:validation} addresses the validation of the extended 
ERPP approach and section \ref{sec:pipe_flow} discusses the simulation setup and parameters for the 
simulations of the turbulent pipe flow, the skin friction coefficient and the mean momentum balance. 
Section \ref{sec:final_remarks} summarises the main findings.


\section{Methodology} \label{sec:methodology}
The flow takes place in the domain ${\cal D}\backslash \Omega$  where $\cal D$ contains fluid and particles.
$\Omega(t) = \cup_p \Omega_p(t)$,  where $\Omega_p(t)$, $p = 1, \ldots N_p$, is the domain occupied by 
the $p$-th particle 
with diameter $d_p$. The fluid is described by the incompressible 
Navier-Stokes equations with the no-slip  condition at the solid boundaries
\begin{equation}
\label{eqn:ns_resolved}
\begin{array}{l}
\left.
\begin{array}{l}
\displaystyle \nabla \cdot \vu = 0 \\ \\
\displaystyle \frac{\partial \vu}{\partial t} + \vu \cdot \nabla \vu = 
-\frac {1}{\rho_f} \nabla {\rm p} + \nu \nabla^2 \vu 
\end{array} 
\right\} \qquad \vx \in {\cal D}\backslash \Omega
\\ \\
\displaystyle \vu\lvert_{\partial \Omega_p(t)} = 
\vV_p(\vx)\lvert_{\partial \Omega_p} \qquad \qquad p =1,\ldots,N_p \\ \\
\displaystyle \vu^{(\pi)}  \lvert_{\partial {\cal D}} = 0 \\ \\
\displaystyle \vu \cdot \vn\lvert_{\partial {\cal D}}  
= u^{(n)}\lvert_{\partial {\cal D}} = 0 
\\ \\
\displaystyle \vu(\vx,0)=\vu_0(\vx) \qquad \qquad \vx 
\in {\cal D}\backslash \Omega^0\ .
\end{array}
\end{equation}
In eq.~(\ref{eqn:ns_resolved}), $\vu_0(\vx)$ is the initial velocity field, $\rho_f$  the fluid 
density, $\nu$  the kinematic viscosity and $\Omega^0 = \Omega(0)$. At the boundaries 
$\partial \Omega_p$  and $\partial {\cal D}$,  impermeability and no-slip conditions are 
assumed. Superscript $n$ and $\pi$ denote normal and tangent components of a given 
vector.

The particles affect the carrier fluid through the no-slip condition at the moving particle 
surface $\partial \Omega_p(t)$ where the fluid matches the local rigid body velocity of the 
particle $\vV_p(\vx) = \vv_p + \vomega_p \times \left(\vx-\vx_p \right)$. The idea is to 
account for the effect of the moving particles on the fluid by defining a suitable correction 
field for which, in the limit of  small particles, a closed form expression can be provided.

Given the current time $t$, for small intervals $t + \tau$, $0 \le \tau \le Dt$, the carrier flow 
velocity is decomposed  into two parts, $\vu(\vx,t+\tau)=\vw+\vv$, that will be referred to as 
the background and perturbation velocity, respectively. The field $\vw(\vx,\tau)$, where 
dependence on the parameter $t$ is dropped for notational simplicity, is  assumed to satisfy 
the equations 
\begin{equation}
\label{eqn:ns_background}
\begin{array}{l}
\begin{array}{l}
\displaystyle \nabla \cdot \vw = 0 \\ \\
\displaystyle \frac{\partial \vw}{\partial \tau} + \vF = 
-\frac {1}{\rho_f} \nabla \pi + \nu \nabla^2 \vw
\end{array}
\\ \\
\displaystyle \vw^{(\pi)} \lvert_{\partial {\cal D}}=-\vv^{(\pi)} \lvert_{\partial {\cal D}} 
\\ \\
\displaystyle \vw \cdot \vn \lvert_{\partial {\cal D}}=w^{(n)}\lvert_{\partial {\cal D}} = 0
\\ \\
\displaystyle \vw(\vx,0)=\vu(\vx,t)\, ,
\end{array}
\end{equation}
where $\vx \in {\cal D}$.
\begin{equation}
\label{eqn:ns_F}
\vF = \left\{
\begin{array}{ll}
\vu \cdot \nabla  \vu & \qquad \mbox{for} \,\, \vx \in   {\cal D}\backslash\Omega\\ \\
\vV_p \cdot \nabla \vV_p & \qquad \mbox{for}  \,\, \vx \in  \Omega
\end{array}
\right. \, , 
\end{equation}
defined in $\cal D$, reproduces the convective term of the Navier-Stokes equation 
in the fluid domain ${\cal D}\backslash \Omega$ and is prolonged inside each particle 
using the corresponding rigid body particle velocity. For the present considerations, 
$\bf F$ can be treated as a prescribed forcing term.  Note that, concerning $\bf w$, 
no boundary conditions are applied to the particle surfaces.

The field $\vv(\vx,\tau)$ exactly satisfies the linear unsteady Stokes equations (the full 
non-linear term being retained in the  equation for $\vw$),
\begin{equation}
\label{eqn:unsteady_stokes}
\begin{array}{l}
\left.
\begin{array}{l}
\displaystyle \nabla \cdot \vv = 0 \\ \\
\displaystyle \frac{\partial \vv}{\partial \tau}
=-\frac{1}{\rho_f} \nabla {\rm q} +\nu \nabla^{2}{\vv}  
\end{array}
\right\} \qquad \vx \in {\cal D}\backslash \Omega
\\ \\
\displaystyle \vv\lvert_{\partial \Omega_p}=\vV_p(\vx)\
\lvert_{\partial \Omega_p} - 
\vw\lvert_{\partial \Omega_p} \qquad p = 1, \ldots N_p 
\\ \\
\displaystyle \vv \cdot \vn \lvert_{\partial {\cal D}}= v^{(n)}\lvert_{\partial {\cal D}} = 0
\\ \\
\displaystyle \frac{\partial \vv^{(\vpi)}}{\partial n}\bigg|_{\partial {\cal D}}= 0
\\ \\
\displaystyle \vv(\vx,0)=0 \qquad \vx \in  {\cal D}\backslash \Omega(\tau)\, .
\end{array}
\end{equation}
The field $\bf v$ is coupled to $\bf w$ through the boundary conditions at the particle surfaces. 
Symmetrically,  $\bf w$ is coupled to $\bf v$ via the external boundary $\partial D$, where 
impermeability and free slip conditions are enforced on $\bf v$. The resulting field $\bf u$ 
satisfies the required impermeability and no-slip conditions at all (particles and external 
domain) solid boundaries. 

Since the linear field $\bf v$ obeys homogenous initial conditions at the initial time 
$\tau = 0$, a simplified integral representation, see e.g. \cite{piva1987vector}, is available for $\vv$,
\begin{equation}
\label{eqn:unsteady_stokes_solution}
v_i(\vx,\tau)=\int_0^\tau d\ttau \int_{\partial \Omega} t_j(\vxi,\ttau) 
\hat{G}_{ij}(\vx,\vxi,\tau,\ttau) 
-v_j(\vxi,\ttau) {\cal \hat{T}}_{ijk} (\vx,\vxi,\tau,\ttau) n_k(\vxi) \, dS_{\vxi},
\end{equation}
where $\hat{G}_{ij}(\vx,\vxi,\tau,\ttau)$ is the Green function, a second order Cartesian tensor, 
appropriate for a free-slip, impermeable, external boundary  $\partial D$. Physically, 
$\hat{G}_{ij}$ is the $i$-th velocity component induced at position $\vx$ and time $\tau$ due 
to a delta function-like impulsive force localised at $\vxi$ acting at time $\ttau$ in direction $j$. 
The stress tensor associated to such velocity field is ${\cal \hat{T}}_{ijk} (\vx,\vxi,\tau,\ttau)$. 
In principle, for a generic domain $\cal D$, the  specific Green function  can be evaluated numerically. 
For the present application, the much simpler free-space solution can be used when the particle 
is far from $\partial D$.  Close to $\partial D$, the actual geometry can be approximated by the 
local tangent plane and the Green function obtained by the method of images.
{This idea consists in describing the effect of the wall through a 
mirrored particle (image particle) that, by superimposing its disturbance flow to the 
flow produced by the physical particle (note that the problem is linear), enforces the
correct boundary condition at the wall, see e.g.~\cite{happel2012low} or~\cite{blake1974fundamental}.}
Equation (\ref{eqn:unsteady_stokes_solution}) expresses $\vv(\vx,\tau)$ in 
terms of a time convolution and a boundary integral involving  the (physical) stress vector 
$t_j(\vxi,\ttau)$ and the perturbation velocity $v_j(\vxi,\ttau)$ at the particle boundaries. 
(No integration on $\partial {\cal D}$ is needed since the domain Green function, or its approximation, is used).

Substituting the first order truncation of the Taylor series of $\hat{G}_{ij}(\vx,\vxi,t,\tau)$ and 
${\cal \hat{T}}_{ijk} (\vx,\vxi,t,\tau)$, centered at the particle geometric centre
$\vx_p$,  in equation (\ref{eqn:unsteady_stokes_solution}) provides the far field disturbance 
velocity, $r_p/d_p \gg 1$, where  $r_p = |\vx -\vx_p |$,
\begin{equation}
\label{eqn:unsteady_stokes_far_field}
v_i(\vx,\tau)=-\sum_p \int_0^\tau D^p_j(\ttau) \hat{G}_{ij}(\vx,\vx_p,\tau,\ttau) \,
d\ttau \ .
\end{equation}
The disturbance field is expressed in terms of the hydrodynamic force  $\vD_p(\tau)$ 
on the particles, with Cartesian components $D^p_j$, and obeys the partial differential equation 
\begin{equation}
\label{eqn:unsteady_stokes_singular}
\begin{array}{l}
\displaystyle \nabla \cdot \vv = 0 \\ \\
\displaystyle \frac {\partial \vv}{\partial \tau} - \nu \nabla^2 \vv + 
\frac{1}{\rho_f} \nabla {\rm q} =  - \frac{1}{\rho_f} \sum_p \vD_p(\tau)
\, \delta\left[ \vx - \vx_p(\tau) \right] 
+ \tilde{\vD}_p(\tau) \, \delta\left[ \vx - \tilde{\vx}_p(\tau) \right] 
\\ \\
\vv(\vx,0)=0 \ . 
\end{array}
\end{equation}
In eq.~\eqref{eqn:unsteady_stokes_singular}, the boundary conditions  on $\partial {\cal D}$ 
are enforced by using the method of images including the additional forcing terms 
{due to the mirrored particles which are indicated by 
the tildes. The image system is obtained by reflection with respect to the local tangent plane 
according to $\tilde{\vx}^{\pi}_p=\vx^{\pi}_p$, $\tilde{x}^{n}_p=-x^{n}_p$, 
$\tilde{\vD}^{\pi}_p=\vD^{\pi}_p$, $\tilde{D}^{n}_p=-D^{n}_p$. The reflection to the local tangent plane
is acceptable when the particle diameter is much smaller than the local curvature of the wall
as will be carefully checked in section \S\ref{sec:validation}}

Following the procedure described in detail in \cite{gualtieri2015exact}, the velocity field obeying 
eq.~\eqref{eqn:unsteady_stokes_singular} can be non-canonically decomposed in the form 
$\vv  = \vv_{\vzeta} + \nabla \phi$,  where the pseudo-velocity $\vv_{\vzeta}$ is the solution of
\begin{equation}
\label{eqn:unsteady_stokes_local}
\begin{array}{l}
\displaystyle \frac {\partial \vv_{\vzeta}}{\partial \tau} - \nu \nabla^2 \vv_{\vzeta} 
 =  - \frac{1}{\rho_f} \sum_p \vD_p(\tau)
\, \delta\left[ \vx - \vx_p(\tau) \right] 
+ \tilde{\vD}_p(\tau) \, \delta\left[ \vx - \tilde{\vx}_p(\tau) \right] 
\\ \\
\vv_{\vzeta}(\vx,0)=0 \ . 
\end{array}
\end{equation}
By taking the curl of eqs.~\eqref{eqn:unsteady_stokes_singular} and \eqref{eqn:unsteady_stokes_local} 
one realises that  $\nabla \times \vv_{\vzeta} = \nabla \times \vv$ with $\nabla \cdot \vv_{\vzeta} \ne 0$.
The complete field $\vv$ is retrieved by projection on solenoidal fields which requires
$\nabla ^2 \phi = - \nabla \cdot \vv_{\vzeta}$. The advantage of this procedure is twofold: i) the 
pseudo-velocity $\vv_{\vzeta}$ is localised around the sources; ii) the correction field $\nabla \phi$ 
can be evaluated a posteriori with the same projection algorithm used to enforce zero divergence of 
the background velocity $\vw$, see \cite{gualtieri2017turbulence} for application to two-way coupled 
particle laden homogeneous shear flows. The field $\vv_{\vzeta}$ can be expressed in  terms of the 
integral representation for the (vector) heat equation, see e.g. \cite{stakgold2000boundary},
\begin{equation}
\label{eqn:pseudo_velo}
\vv_{\vzeta}(\vx,\tau) = - \frac{1}{\rho_f} \int_0^{\tau^+} 
\vD_p(\ttau) g\left[ \vx-\vx_p(\ttau),\tau-\ttau\right]  +
\tilde{\vD}_p(\ttau) g\left[ \vx-\tilde{\vx}_p(\ttau),\tau-\ttau\right]  
\, d\ttau 
\end{equation}
where the method of images has been used as before to enforce the boundary conditions on 
$\partial D$ and the free space Green's function reads
\begin{equation}
\label{eqn:fundamental_fourier}
g(\vx-\vxi,\tau-\ttau) = \frac{1}{\left[ 4 \pi \, \nu (\tau-\ttau)\right]^{3/2}}
\exp\left[-\frac{\lVert \vx -\vxi \rVert ^2}{4 \nu (\tau-\ttau)} \right] \ .
\end{equation}
$\vv_{\vzeta}$ is a singular field that can be regularised by limiting the
upper integration limit to $\tau-\varepsilon_R$, with $\varepsilon_R << 1$ a small regularisation 
parameter. The partial differential equation for the regularised field turns out to be, 
\begin{equation}
\label{eqn:v_zeta_regular}
\begin{array}{ll}
\displaystyle \frac{\partial \vv_{{\vzeta}_R}}{\partial \tau} - \nu \nabla^2 \vv_{{\vzeta}_R}  = 
- \frac{1}{\rho_f} 
& \left\{ \vD_p(\tau-\epsilon_R) g\left[\vx - \vx_p(\tau-\epsilon_R), \epsilon_R \right] +
\right.  \\
&+ \left. \tilde{\vD}_p(\tau-\epsilon_R) g\left[\vx - \tilde{\vx}_p(\tau-\epsilon_R),
\epsilon_R \right] \right\}  \, ,
\end{array}
\end{equation}
where again the boundary conditions are taken int account through the method of images. It is noteworthy 
that the forcing field is now expressed as a collection of Gaussians with small but finite variance 
($\sigma(\tau) = \sqrt{2 \nu (\tau - \varepsilon_R)} \ge \sigma_R = \sqrt{2 \nu \varepsilon_R}$) that 
can be discretised on a finite grid, provided the grid size is smaller than the minimum variance 
($Dx < \sigma_R$). {Note that the effect of the image particle decays in space
faster than exponentially, hence its contribution may be neglected when the particle distance from the
walls equals a few variances, say $3 \sigma_R$}
Another crucial aspect to take into account is the time delay in the position 
and drag of the particles, evaluated at the earlier time instant $\tau - \varepsilon_R$.
The regularisation procedure amounts to removing the effect of the vorticity generated by the drag force 
exerted by the particle on the fluid in the last time instants $\ttau \ge \tau - \varepsilon_R$, eq.~\eqref{eqn:pseudo_velo}. 
Such singular vorticity field cannot be resolved by a finite grid. It is not however neglected, since it is taken into account 
at later times, after it is spread out by diffusion. This aspect is of paramount importance to guarantee exact momentum 
conservation in the particle-fluid interaction and prevent the incurred error to accumulate in time, see \cite{gualtieri2015exact}. 

The two fields, $\vw$ and $\vv_R = \vv_{\vzeta_R} + \nabla \phi_R$ ($\phi_R$ being the 
potential correction needed to make $\vv_R$ solenoidal) can now be recombined in the complete, 
regularised velocity $\vu_R = \vw + \vv_R$, whose evolution equation is
\begin{equation}
\label{eqn:ns_regularized_filtered}
\begin{array}{ll}
\displaystyle \nabla \cdot \vu_R = 0 \\ \\
\displaystyle \frac{\partial \vu_R}{\partial \tau} + \vu_R \cdot \nabla \vu_R
 =  -\frac {1}{\rho_f} \nabla p + \nu \nabla^2 \vu_R 
- \frac{1}{\rho_f} \sum_p^{N_p} 
&\left\{ \vD_p(\tau-\epsilon_R) \, g\left[ \vx-\vx_p(\tau-\epsilon_R),\epsilon_R \right] \right. + \\
& \left. \tilde{\vD}_p(\tau-\epsilon_R) \, g\left[ \vx-\tilde{\vx}_p(\tau-\epsilon_R),\epsilon_R \right]  \right\} \\ \\
\displaystyle \vu^{(\vpi)}_R\lvert_{\partial {\cal D}} = 0 &   \\ \\
\displaystyle \vu \cdot \vn = u^{(n)}_R\lvert_{\partial {\cal D}} = 0 &   \\ \\
\displaystyle \vu_R(\vx,0)=\vu(\vx,t)  \ .&    
\end{array}
\end{equation}
Finally, the no-slip condition on $\partial {\cal D}$ 
in presence of the perturbation induced by particles is worth discussing.
The background field $\vw$ can be interpreted as the superposition of 
two other fields, $\vw=\bar{\vw}+\vw^\prime$. $\bar{\vw}$, satisfying the Navier-Stokes equations where
the standard advection term is replaced by $\vF$, as in \eqref{eqn:ns_F}, with
impermeability and no-slip at $\partial {\cal D}$ and initial condition $\bar{\vw}(\vx,0)=\vu(\vx,t)$.
Since the full non linear term is accounted for by $\bar{\vw}$,  $\vw'$ satifies the unsteady Stokes equations
\begin{equation}
\label{eqn:first_stokes_problem}
\begin{array}{l}
\displaystyle \nabla \cdot \vw^\prime = 0 \\ \\
\displaystyle \frac{\partial \vw^\prime}{\partial \tau} = 
-\frac {1}{\rho_f} \nabla \pi^\prime + \nu \nabla^2 \vw^\prime
\\ \\
\displaystyle \vw^\prime_{(\vpi)} \lvert_{\partial {\cal D}}=-\vv^{(\vpi)}_R \lvert_{\partial {\cal D}}
\\ \\
\displaystyle \vw^\prime \cdot \vn \lvert_{\partial {\cal D}}=w^\prime_{(n)}\lvert_{\partial {\cal D}} = 0
\\ \\
\displaystyle \vw^\prime(\vx,0)=0 \, , 
\end{array}
\end{equation}
where the slippage imposed on $\partial {\cal D}$ balances the slip velocity due to the particle disturbance field.
This is a generalisation of the well-known Stokes first problem for a flat plate which 
starts moving impulsively from rest. As in this classical problem, the slip velocity at the wall can be interpreted
as a vortex sheet which is subsequently diffused in the flow domain, \citep{benfatto1984generation},
mimicking the mechanism of vorticity generation at the wall, see \cite{morton1984generation} and \cite{casciola1996vorticity}.

\section{Validation} \label{sec:validation}
The method is validated by considering the global impulse 
balance. In free space, the coupling algorithm was already shown to conserve 
total momentum in \cite{gualtieri2015exact}. The conservation properties of 
the extended algorithm in presence of a solid wall are now discussed. 
The simple but stringent tests carried out are instrumental to  
turbulent wall-bounded flows where the particles are known to accumulate
in the near wall region, making momentum exchange between particles, fluid and 
the solid wall crucial. 

A basic test case considers the fluid motion induced by a constant 
force, $\vF$, applied at a fixed point, $\vx_p$, to the fluid initially at rest 
in presence of solid boundaries. A cylindrical domain ${\cal D}$, of 
circular cross-section with radius $R$, is considered and the field is assumed 
to be periodic in the axial direction $z$. In cylindrical coordinates,  
$\vx_p=(r_p,\theta_p,z_p)$, the applied force and the velocity field read
$\vF=(F_r, F_\theta, F_z)$ and  $\vu=(u_r, u_\theta, u_z)$, respectively.  
The radial wall-normal distance is denoted by $y_p=R-r_p$. The constant force 
is applied in the $z$-direction, $\vF=(0,0,F_0)$, and the impulse grows linearly 
in time, $\vI=\vF \, t$. 

\begin{figure}
\centering{
\includegraphics[width=.48\textwidth]{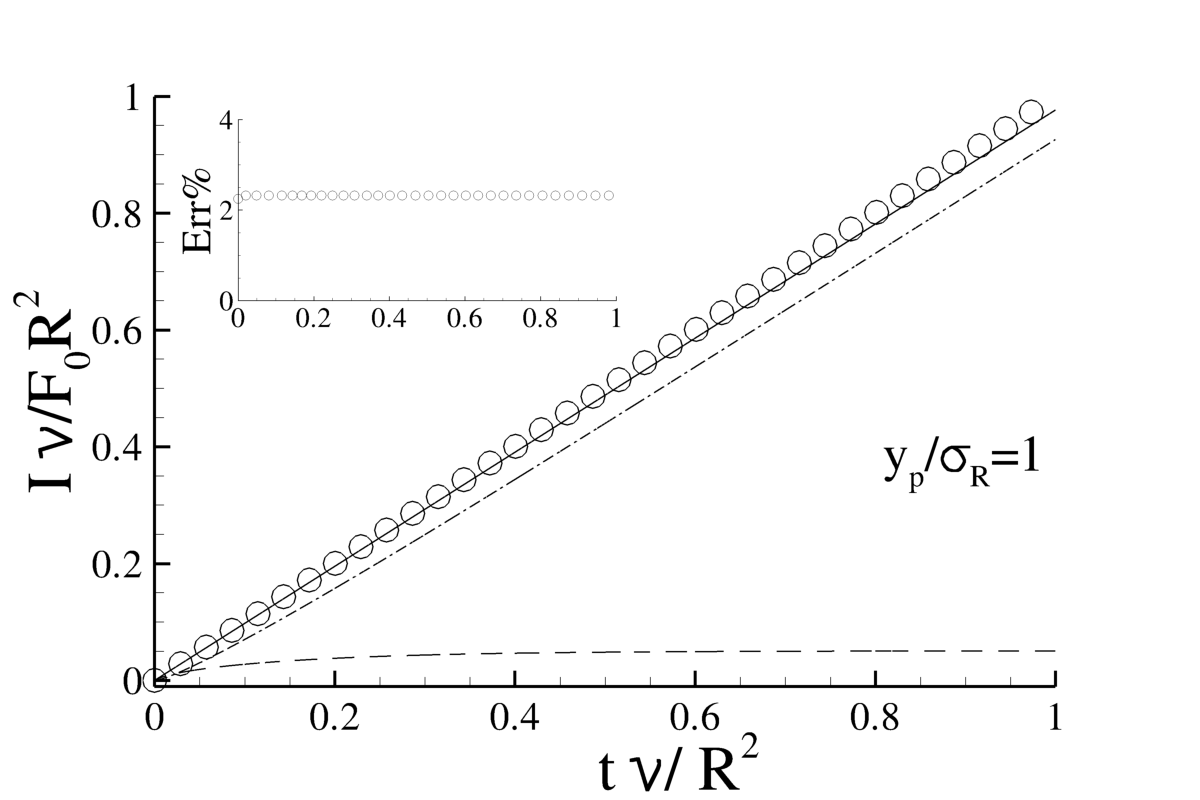}
\includegraphics[width=.48\textwidth]{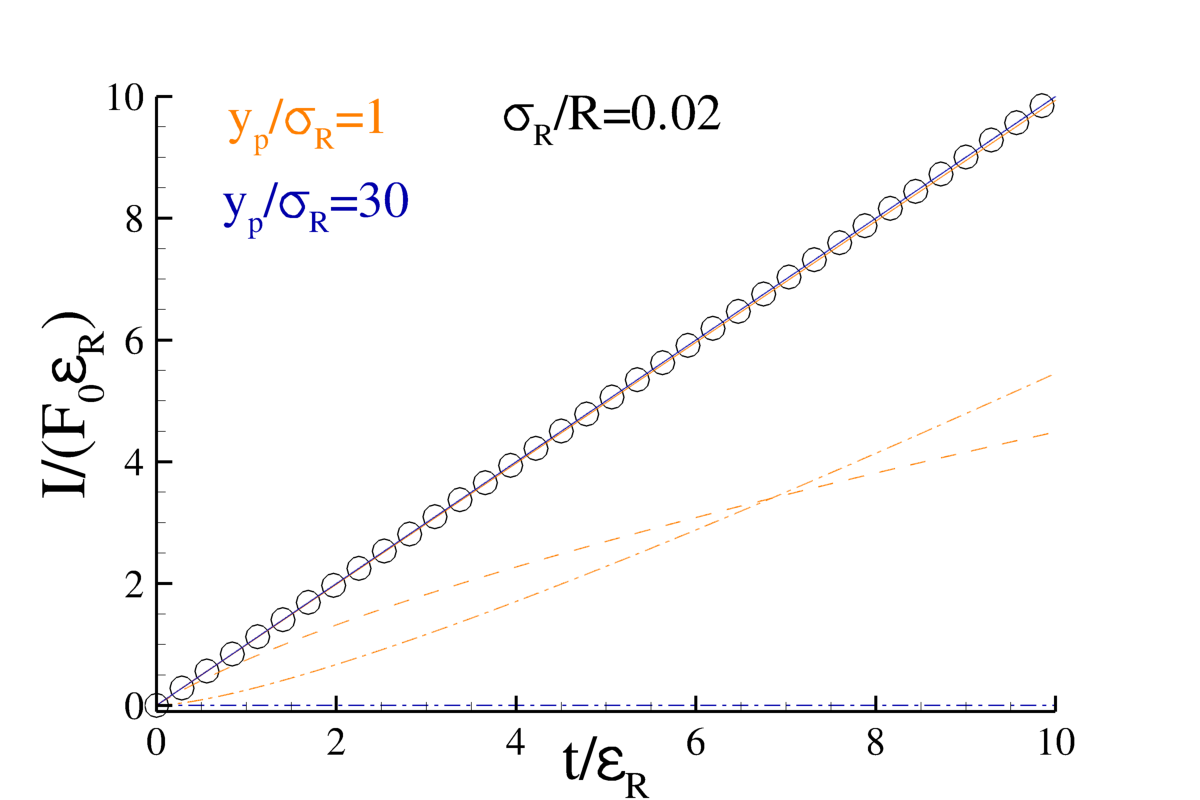}
{\scriptsize \put(-370,110){\bf (a)}}
{\scriptsize \put(-185,110){\bf (b)}}
\includegraphics[width=.48\textwidth]{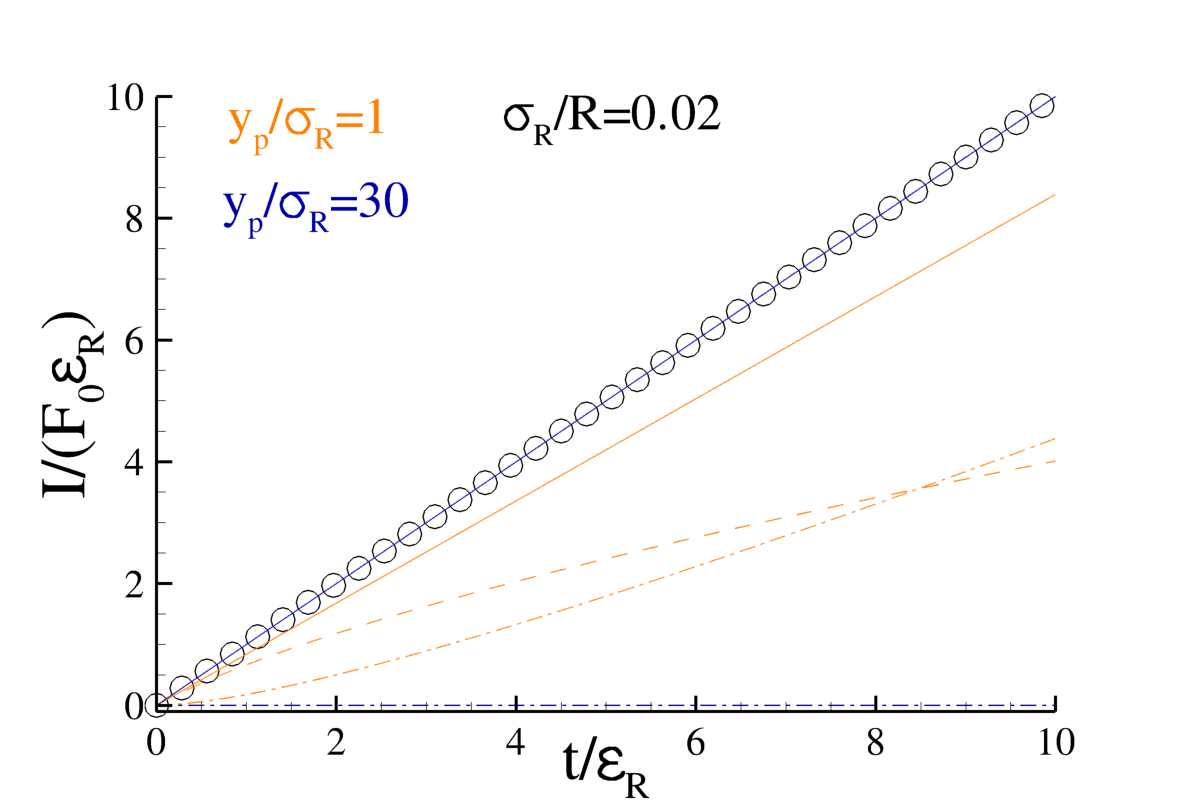}
\includegraphics[width=.48\textwidth]{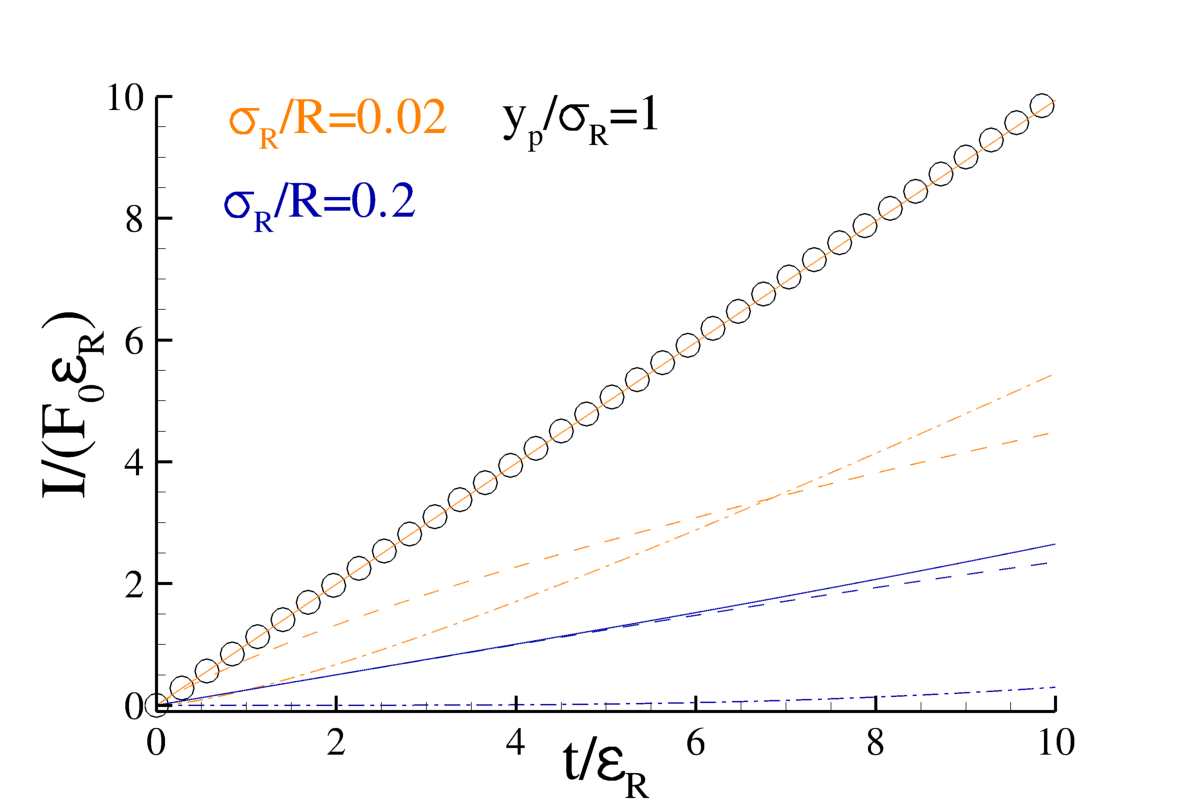}
{\scriptsize \put(-370,110){\bf (c)}}
{\scriptsize \put(-185,110){\bf (d)}}
}
\caption{\label{fig:impulse_single}
Impulse balance (\ref{eqn:impulse_balance}) for a constant 
force $\vF = (0,0,F_0)$ applied at a fixed point $\vx_p$ at distance
$y_p=R-r_p$ from the wall. The fluid is initially at rest in the cylindrical 
domain ${\cal D}$ with periodic boundary condition at $z=0$ and $z=2\pi$ and 
no-slip boundary conditions at $r=R$. Exact impulse $I_E=F_0 \,t$ (open circles), 
$I_u(t)$ (dashed line), $I_f(t)$ (dash-doted line) and $I_u + I_f$ (solid line). 
Panel a): plot of the impulse normalised in external variables 
$\nu I/\left(F_0 \, R^2 \right)$ versus dimensionless time
$\nu t /R^2$. The inset shows the normalised relative error 
$Err\% =100 \, \cdot (I_E-I_u-I_f)/I_E$.
Panels b)-d): plot of the impulse normalised in internal variables 
$I/\left(F_0 \, \epsilon_R \right)$ versus dimensionless time
$t/\epsilon_R$. 
Panels b) and c): colours label different wall normal distances 
made dimensionless with the regularisation length-scale $\sigma_R$, 
namely $y_p/\sigma_R=1$ (orange lines) and $y_p/\sigma_R=30$ 
(blue lines) at $\sigma_R/R=0.02$. Panel b): cases with image point-force.
Panel c): same cases as in panel b) without the image point-force.
Panel d): the colours label the different regularisation length-scale $\sigma_R$
made dimensionless with the pipe radius $R$, namely $\sigma_R/R=0.02$ (orange lines)
and $\sigma_R/R=0.2$ at for a point-force at fixed $y_p/\sigma_R=1$.
}
\end{figure}

Time integration of the global axial force balance, 
${\partial  I_u}/{\partial t} = D_f + F_0$, 
where $I_u(t)=\int_{{\cal D}} \rho_f u_z \, dV$ and 
$D_f=\int_{\partial {\cal D}} \mu \,{\partial u_z}/{\partial r} \, dS$ are
the fluid impulse and the viscous drag force, respectively, yields 
\begin{equation}
\label{eqn:impulse_balance}
I_u(t) + I_f(t) = F_0 \, t \,
\end{equation}
where $I_f(t)=-\int_0^t D_f(\tau) d\tau$ is the impulse due to friction drag.
In dimensionless form, the different terms take the form 
$\nu I_u/(F_0  R^2) = I^*_u (\nu t/R^2)$. In the ERPP method, the new parameter 
$\nu \varepsilon_R/R^2$ appears associated with the regularisation time scale 
$\varepsilon_R$, e.g. 
$\nu I_u/(F_0  R^2) = I^*_u (\nu t/R^2, y_p/R, \nu \varepsilon_R/R^2)$,
\begin{equation}
\label{eqn:impulse_balance_adim_ext}
I_u^*\left(\frac{\nu t}{R^2},\frac{y_p}{R},\frac{\nu\varepsilon_R}{R^2}\right) 
+I_f^*\left(\frac{\nu t}{R^2},\frac{y_p}{R},\frac{\nu \varepsilon_R}{R^2}\right) 
= \frac{\nu t}{R^2} \, ,
\end{equation}
where the original form \eqref{eqn:impulse_balance} is recovered in the limit 
$\nu \varepsilon_R/R^2$ approaching zero. 
Figure~\ref{fig:impulse_single}(a) corresponds to a test case with 
the force applied close to the wall ($y_p/R = 0.1$) and  
$\nu \varepsilon_R/R^2 = 4 \cdot 10^{-3}$, and shows that the impulse balance is 
satisfied within numerical accuracy on the (external) time scale $R^2/\nu$. 
At steady state, the fluid impulse becomes constant and the drag impulse increases 
linearly with time, becoming dominant at large time. 
{The correct evaluation of the friction drag impulse appears now
in all its relevance for wall-bounded flows. Indeed,  the approach we propose 
is able to generate vorticity at the wall in a physically consistent way as proved
by the correct evaluation of the viscous shear stress at the wall.}
The relative error between
the exact value of the total impulse $I_E$ and its numerical evaluation 
$I_u+I_f$ is shown in the inset of panel a). The error does not
accumulate in time.

A more subtle test concerns the impulse balance on the (inner) time scale of the 
regularisation parameter. Using $\varepsilon_R$ as time in the dimensional 
analysis yields 
\begin{equation}
\label{eqn:impulse_balance_adim_int}
I_u^{**}\left(\frac{t}{\varepsilon_R}, \frac{y_p}{\sqrt{2 \nu \varepsilon_R}},
\frac{R}{\sqrt{2 \nu \varepsilon_R}}\right) + 
I_f^{**}\left(\frac{t}{\varepsilon_R}, \frac{y_p}{\sqrt{2 \nu \varepsilon_R}},
\frac{R}{\sqrt{2 \nu \varepsilon_R}}\right) = \frac{t}{\varepsilon_R} \, , 
\end{equation}
where, e.g., $I_u/(F_0 \varepsilon_R) = I_u^{**}$ and, as in 
\S~\ref{sec:methodology}, $\sqrt{2 \nu \varepsilon_R} = \sigma_R$. 
This alternative dimensionless form stresses the behaviour of the solution on 
the time scale of the regularisation, corresponding to the diffusive length 
scale $\sigma_R$, which is of the order of the mesh size to be adopted in the 
numerical solution. The purpose of checking the impulse balance in the above form 
is a more stringent check of the boundary conditions. In the 
theoretical description of the approach, the Green's function of the domain 
was approximated using the method of images, mirroring the source with respect 
to the local tangent plane at the boundary. 

Figure~\ref{fig:impulse_single}(b) shows the impulse balance for two 
wall-normal distances of the point force. In one case $y_p / \sigma_R=1$ 
(orange solid line)  the distance of the source from the wall is comparable to  
the regularisation length scale $\sigma_R$. In the other, $y_p / \sigma_R=30$ 
(blue solid line, almost totally superimposed on the orange one),  the point force is 
relatively far from the wall. In both cases, the numerically evaluated impulse 
follows the exact solution (circles). In the first case (particle close to the wall) 
the fluid impulse, $I_u$ (orange dashed line) is initially comparable with the drag 
impulse $I_f$ (orange dash-dotted line). On the contrary, in the second case, when 
the force is applied far from the boundary, the friction drag is negligible on 
the observed (inner) time scale and the total impulse is almost all provided by the 
fluid. Overall, the result shows that the boundary condition and the associated 
vorticity generation is correctly captured by the algorithm. Panel c) 
illustrates the role of mirror image of the force by plotting results obtained 
by removing the image contribution. One expects that the effect of the image 
should be negligible when $y_p/\sigma_R \gg 1$. This is indeed the case, as 
shown by the comparison of the blue lines with the corresponding ones in panel b).
On the contrary, when the distance of the application point is comparable with 
the regularisation length, $y_p/\sigma_R \le 1$, the contribution of the image 
is crucial, as seen when comparing the orange solid lines with the corresponding ones 
in panel b). Since, for computational efficiency, the adopted Green's function is 
only approximate, it is important to identify the range of validity of the 
approximation. The curvature of the wall, measured in terms of the regularisation 
length scale, is crucial parameter that determines the accuracy. Panel d) shows 
that, when $\sigma_R/R$ is sufficiently small (orange curves), the error is 
negligible. The error becomes larger as soon as this ratio increases (blue 
curves, wall curvature comparable with the regularisation length).

\begin{figure}
\centering{
\includegraphics[width=.48\textwidth]{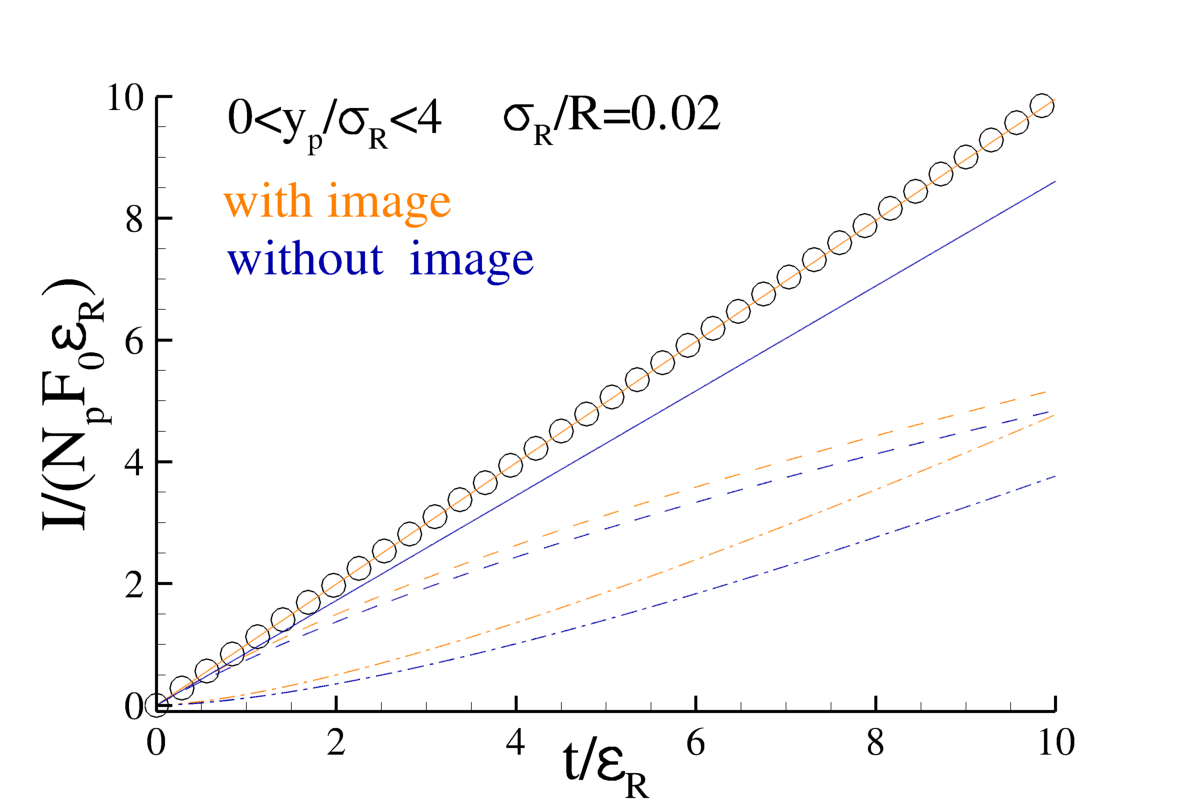}
}
\caption{\label{fig:impulse_many}
Impulse balance (\ref{eqn:impulse_balance}) for a constant 
force $\vF = (0,0,F_0)$ applied to $N_p$ fixed points $\vx_p$ in
the cylindrical domain ${\cal D}$ to the fluid initially 
at rest. The normalised impulse $I/\left(N_p \, F_0 \, \epsilon_R \right)$ 
is plotted versus time $t$ made dimensionless with the regularisation time-scale 
$\epsilon_R$. Exact impulse $I_E$ (open circles), $I_u(t)$ (dashed line), 
$I_f(t)$ (dash-doted line) and $I_u + I_f$ (solid line).
$N_p=100000$ point forces applied at points $\vx_p$ 
randomly distributed in the stripe $0 < y_p/\sigma_R <4$ near the wall for $\sigma_R/R=0.02$.
The colours label cases with image point forces (orange lines)
and without the images (blue line).
}
\end{figure}

Figure~\ref{fig:impulse_many} stresses the results of the previous figure, with 
emphasis on turbulent wall-bounded flows. As discussed in more detail in 
the following sections, inertial particles tend to accumulate in the viscous 
sublayer near the wall. The data reported in the figure artificially reproduce 
these conditions, by considering $N_p=100000$ randomly distributed point forces 
placed in an annular shell close to the cylindrical wall. As apparent in the 
plots, using the mirror images (orange curves) provides the total impulse. On the 
contrary, neglecting the images (blue symbols) completely spoils the quality 
of the simulation. 

\section{Particle-laden turbulent pipe flow} \label{sec:pipe_flow}

\subsection{Simulation setup} \label{sec:sim_setup}

The ERPP formulation is applied to  a fully developed turbulent pipe flow.
The dimensionless forms of equations \eqref{eqn:ns_regularized_filtered}  
are solved in a cylindrical domain 
${\cal D} = [0:R] \times [0: 2\pi] \times [0:L_z] $
where the (dimensionless) pipe radius is $R = 1$. Periodic boundary conditions are 
applied in the axial ($z$) direction, with $L_z = 2 \pi$.
The subscript $R$ which is used to denote the regularised field will be 
dropped hereafter to ease notation. The reference quantities are the fluid 
density $\rho_f$, the pipe radius $R$, the bulk velocity of the purely 
Newtonian case $U_b = Q_0/(\pi R^2)$, where $Q_0$ is the flow rate of the 
reference uncoupled case, and the viscosity $\mu$.

The flow is sustained by a constant mean pressure gradient applied in the  
direction of the axial unit vector $\ve_z$, with the dimensionless pressure 
expressed as $P = dp/dz|_0 (z-z_0) + p(r,\theta,z,t)$, 
\begin{equation}
\label{eqn:ns_pipe}
\begin{array}{ll}
\displaystyle \nabla \cdot \vu = 0 \\ \\
\displaystyle \frac{\partial \vu}{\partial t} +  \nabla \cdot \left(\vu \otimes \vu\right)  
= - \nabla p + \frac{1}{{\rm Re}_b} \nabla^2 \vu + \vf
- \frac{dp}{dz}\Big|_0 \ve_z \ .
\end{array}
\end{equation}
The Reynolds number is ${\rm Re}_b = U_b R/\nu$ and the field $\vf$ is the 
particle feedback on the fluid 
\begin{equation}
\label{eqn:def_back_reaction}
\vf = -\sum_p^{N_p} 
\vD_p(t-\epsilon) \, g\left[ \vx-\vx_p(t-\epsilon),\epsilon\right] 
+\tilde{\vD}_p(t-\epsilon) \, g\left[ \vx-\tilde{\vx}_p(t-\epsilon),
\epsilon \right]  \ .
\end{equation}
The system consists of the carrier Newtonian fluid and of $N_p$ particles. 
The dimensionless drag force on the $p$-th particle is 
${\vD}_p = 3 \pi d_p/{\rm Re_b} \left(\vu|_p + d_p^2/24 \nabla^2 \vu|_p - \vv_p \right)$, 
where $d_p$ is the dimensionless particle diameter, $\vv_p$ the particle velocity and 
$\vu|_p = \vu(\vx_p,t)$ is the fluid velocity at the particle position. Both the 
current time $t$ and the regularisation time scale $\epsilon_R$ are made 
dimensionless with $R/U_b$, that is $\epsilon = \epsilon_R U_b/R$.

\begin{table}
\begin{centering}
\begin{tabular}{cccccc}
  $\phi$ & $\rho_p/\rho_f$ & $St_+$ &$St_0$ & $d_p^+$ & $N_p$   \\ \\ \hline \hline
  0      & -               & -      & -     & -       & -       \\ \hline
  0.1    & 180             & 10     & 0.82  & 1       & 122145 \\
  0.2    & 180             & 10     & 0.82  & 1       & 244290 \\
  0.4    & 180             & 10     & 0.82  & 1       & 488580 \\
  0.6    & 180             & 10     & 0.82  & 1       & 732870 \\ \hline
  0.4    & 180             & 15     & 1.23  & 1.23    & 265950 \\
  0.4    & 180             & 20     & 1.64  & 1.41    & 172739 \\
  0.4    & 180             & 80     & 6.54  & 2.82    & 21592 \\\hline
  0.4    & 90              & 10     & 0.82  & 1.41    & 345479 \\ 
  0.4    & 360             & 10     & 0.82  & 0.70    & 690957 \\
  0.4    & 560             & 10     & 0.82  & 0.57    & 861775 \\
  0.4    & 900             & 10     & 0.82  & 0.45    & 1092499 \\ \hline
\end{tabular}
\caption{Simulation matrix. All runs are performed by imposing the same mean 
pressure gradient corresponding to a friction Reynolds number of $Re_*=180$. 
The bulk Reynolds number for the reference case where the particles do not 
back-react on the fluid (one-way coupling) is $Re_b=2650$. The
grid resolution is $N_\theta \times N_r \times N_z = 576 \times 129 \times 576$ 
corresponding to $\Delta r^+\big|_w = 0.5$ at the wall and
$\Delta r^+\big|_0=2$ at the centerline. The resolution in the azimuthal
and axial directions is  $(R \Delta \theta)_+=3.2$ and $\Delta z_+=3.2$, 
respectively. The mass loading is defined as $\phi=\rho_p N_p V_p/ \rho_f V_f$ 
where $N_p$ is the number of particles, $V_f$ is the volume of the fluid in the 
domain ${\cal D} = [0:R] \times [0: 2\pi] \times [0:L_z] $ and $\rho_p/\rho_f$ 
denotes the particle-to-fluid density ratio. $St_+$ is the Stokes number in 
internal units and $St_0$ is the Stokes number in external units, namely 
$St_0=\tau_p/\tau_0$ with $\tau_0=R/U_b$ being $U_b$, the bulk velocity in the 
uncoupled case. The column labeled $d_p^+$ shows the particle diameter in wall 
units. $N_p$ denotes the number of particles in the domain ${\cal D}$.
\label{tab:sim_matrix}}
\end{centering}
\end{table}

Impermeability and no-slip conditions are enforced at the pipe wall. 
Equations \eqref{eqn:ns_pipe} are solved in cylindrical coordinates by
exploiting a second order finite difference on a staggered grid,
see~\cite{costantini2018drag,battista2014turbulent}. The classical Chorin's projection method, 
\cite{chorin1968numerical,rannacher1992chorin}, is used to enforce the 
divergence-free constraint 
imposed by the mass balance. Both convective and diffusive terms are explicitly 
integrated in time using a third-order low-storage Runge-Kutta method.

As customary, inner or wall units are given in terms of the viscous length 
$\ell_*=\nu/u_*$ and the friction velocity $u_*=\sqrt{\tau_w/\rho_f}$, with  
$\tau_w$ the average wall shear stress. The distance from the pipe wall in inner 
units is denoted $y_+ = (1-r) {\rm Re_*}$, where the friction Reynolds number is
${\rm Re}_* = u_* R / \nu$. The same distance in external units is denoted by 
$y = 1-r$.

All the simulations are performed with the same friction Reynolds number 
$Re_*=180$. 
The corresponding bulk Reynolds number for a purely Newtonian (no particle 
backreaction) flow is $Re_b=2650$.  The grid resolution is 
$N_\theta \times N_r \times N_z = 576 \times 129 \times 576$ in the 
azimuthal, wall-normal and axial direction respectively. The grid in the radial direction is 
clustered near the wall with a minimum spacing of $\Delta r_+\big|_w = 0.5$ which 
gradually increases towards the centreline reaching $\Delta r_+\big|_0=2$. The grid
resolution in the azimuthal and axial direction is $(R \Delta \theta)_+=3.2$
and $\Delta z_+=3.2$ respectively.

Given the large particle-to-fluid density ratio $\rho_p/\rho_f$, the only relevant 
hydrodynamic force is the Stokes drag where the Faxen correction is accounted for, 
see \cite{maxril,gatignol}. The Newton's equations for the particles reduce to 
\begin{equation}
\label{eqn:eq_particles}
\begin{array}{ll}
\displaystyle \frac{d\vx_p}{dt}= \vv_p \\ \\
\displaystyle  \frac{d \vv_p}{dt}= \frac{1}{{\rm St}_b}
\left(\vu|_p +\frac{d_p^2}{24} \nabla^2 \vu|_p - \vv_p \right) \, ,
\end{array}
\end{equation}
where the bulk Stokes number is 
${\rm St}_b = \tau_p U_b/R = \rho_p/ (18 \rho_f) {\rm Re}_b d_p^2$, with
$\tau_p$ the Stokes relaxation time of the particle. 

In eqs.~\eqref{eqn:eq_particles} and in the expression for the drag force 
\eqref{eqn:def_back_reaction},  $\vu|_p$ and $\nabla^2 \vu|_p$ are the fluid 
velocity and its Laplacian evaluated at the particle position taking into 
account the background fluid 
velocity including the disturbance of all the particles except the $p$-th one. 
This field is evaluated by summing the contributions of all the 
particles and by successively removing the particles' self-disturbance.
This step is easily performed with the ERPP approach where
the self-disturbance velocity can be computed in a closed analytical form.  
\begin{figure}
\centering{
\includegraphics[width=.95\textwidth]{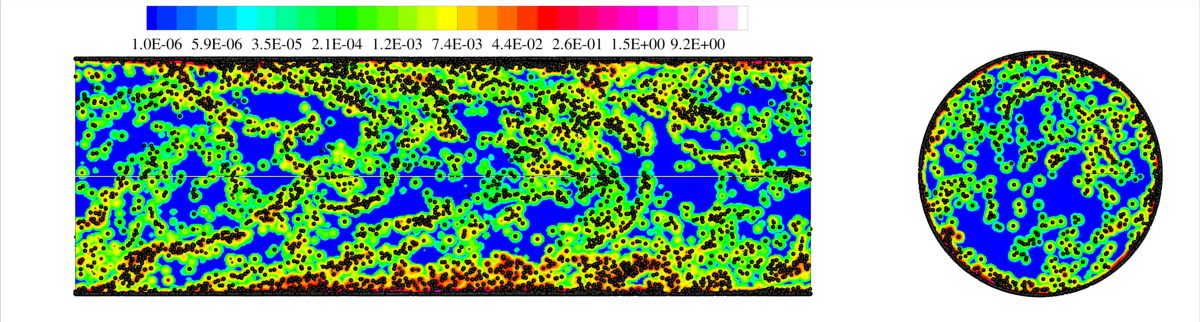}
}
\caption{Snapshot of the particle back reaction field intensity (color) and particle configuration 
(black dots) for an instantaneous field at $\phi=0.4$, $St_+=10$ and $\rho_p/\rho_f=180$. 
The flow is from left to right.}
\label{fig:inst_feedback}
\end{figure}

\begin{figure}
\centering{
\includegraphics[width=.70\textwidth]{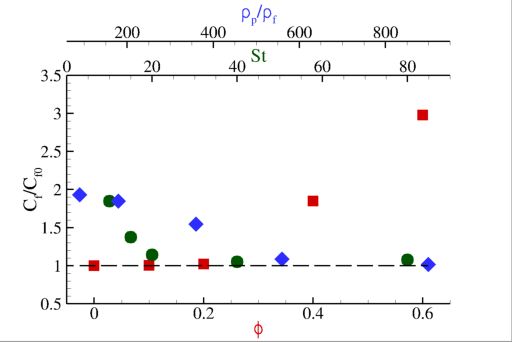}
}
\caption{Friction coefficient of the two-way coupled simulations $C_f$ 
normalised with corresponding friction coefficient of the uncoupled case 
$C_{f0}$, see eq.~\eqref{eqn:fric_coef}. The dataset is plotted as a 
function of the the mass loading $\phi$ (red squares) at fixed $St_+=10$ and 
$\rho_p/\rho_f=180$, as a function of the Stokes number $St_+$ 
(green circles) at fixed $\phi=0.4$ and $\rho_p/\rho_f=180$, and as a 
function of the particle to fluid density ratio $\rho_p/\rho_f$ 
(blue diamonds) at fixed $\phi=0.4$ and $St_+=10$. }
\label{fig:fric_coef}
\end{figure}

It is instrumental to introduce the inner Stokes number,
${\rm St}_+=\tau_p\ell_*/u_* = {\rm St}_0 {\rm Re}_*^2/{\rm Re}_b $. 
In two-way coupled simulations, a further dimensionless parameter that quantifies  
the particle backreaction on the fluid is the mass loading of the 
suspension. This is defined as the ratio between the total mass of the disperse phase 
and the fluid mass, 
$\phi=N_p \rho_p V_p / \rho_f V_f = (\rho_p / \rho_f) N_p  d_p^2 /(12 \pi)$, where 
$V_p$ is the volume of the particle and $V_f = \pi R^3 L_z$ ($L_z = 2 \pi$ is 
the dimensionless axial extension of the domain) is the volume of the 
fluid in the domain ${\cal D}$. In the expression for the mass loading,  
$\phi_V=N_p V_p /V_f = N_p d_p^2/(12 \pi)$ is the volume fraction. 

In conclusion, the dynamics is controlled by a set of four dimensionless 
parameters, $\left\{ Re_*; \, St_+; \, \phi; \, \rho_p/\rho_f \right\}$. 
The physical assumptions behind this description of the particle laden flow are: 
i) the density ratio $\rho_p/\rho_f$ is sufficiently large such that only the 
Stokes drag matters in the particle dynamics and 
ii) the particle diameter $d_p^+ = d_p Re_*$ is small, which means that the 
particles are at most of the same order of magnitude of the viscous length. 

The parameters for the different cases are summarised
in table~\ref{tab:sim_matrix}. The friction Reynolds number is fixed 
(i.e. the pressure drop is constant). The simulations are divided into 
three groups.  In the first set, the mass loading $\phi$ is changed keeping  
Stokes number and density ratio fixed. The second set addresses the 
effects of the Stokes number at fixed mass loading and density ratio. Finally, 
the density ratio is changed at fixed mass loading and Stokes number to explore 
the effect of the number of particles.

{A snapshot of the particle back-reaction intensity on the fluid and the instantaneous 
particle configuration is provided in figure \ref{fig:inst_feedback} for the reference
case at $\phi=0.4$, $St_+=10$ and $\rho_p/\rho_f=180$. Note the particle accumulation in the
near-wall region and the strict correlation between the particle configuration and the Eulerian
structure of the back reaction. Coherent particle structures extend from the wall up to the
center of the pipe resembling the hairpin-like structures typical of wall-bounded flows.}

\subsection{Skin friction coefficient} \label{sec:skin_fric}
\begin{figure}
\centering{
\includegraphics[width=.48\textwidth]{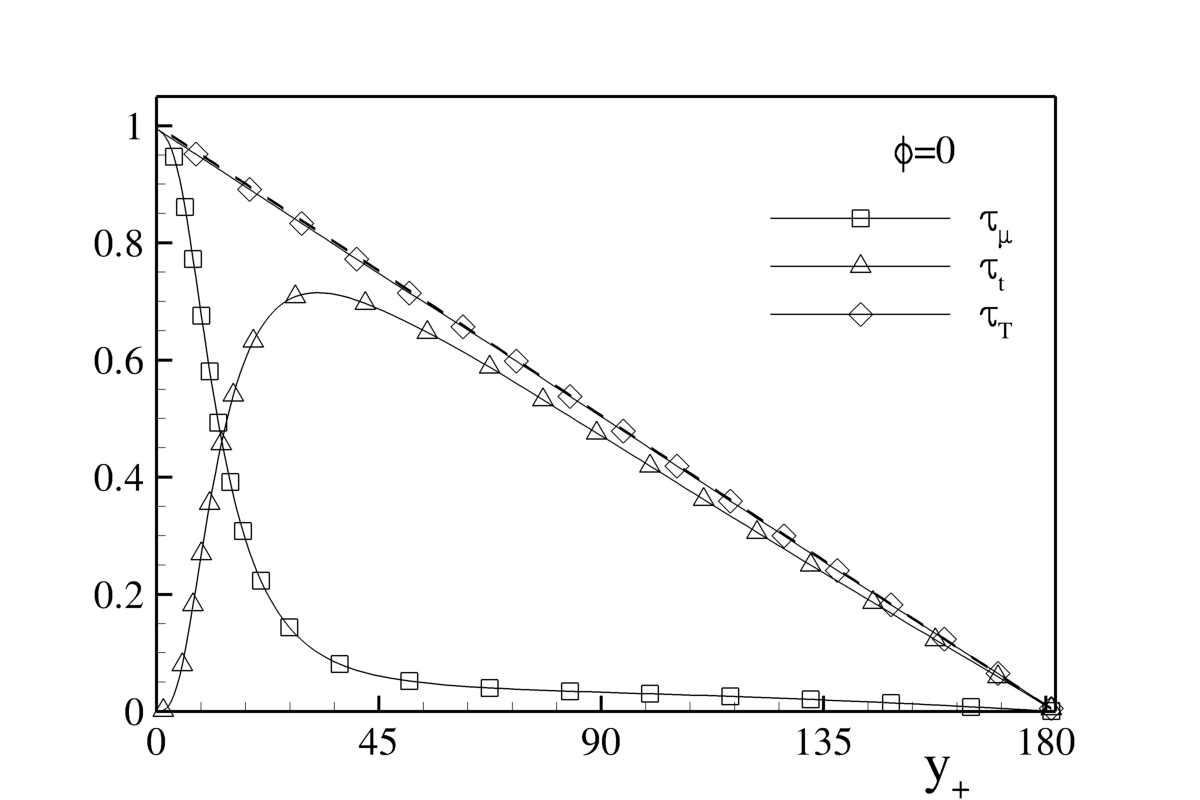}
\includegraphics[width=.48\textwidth]{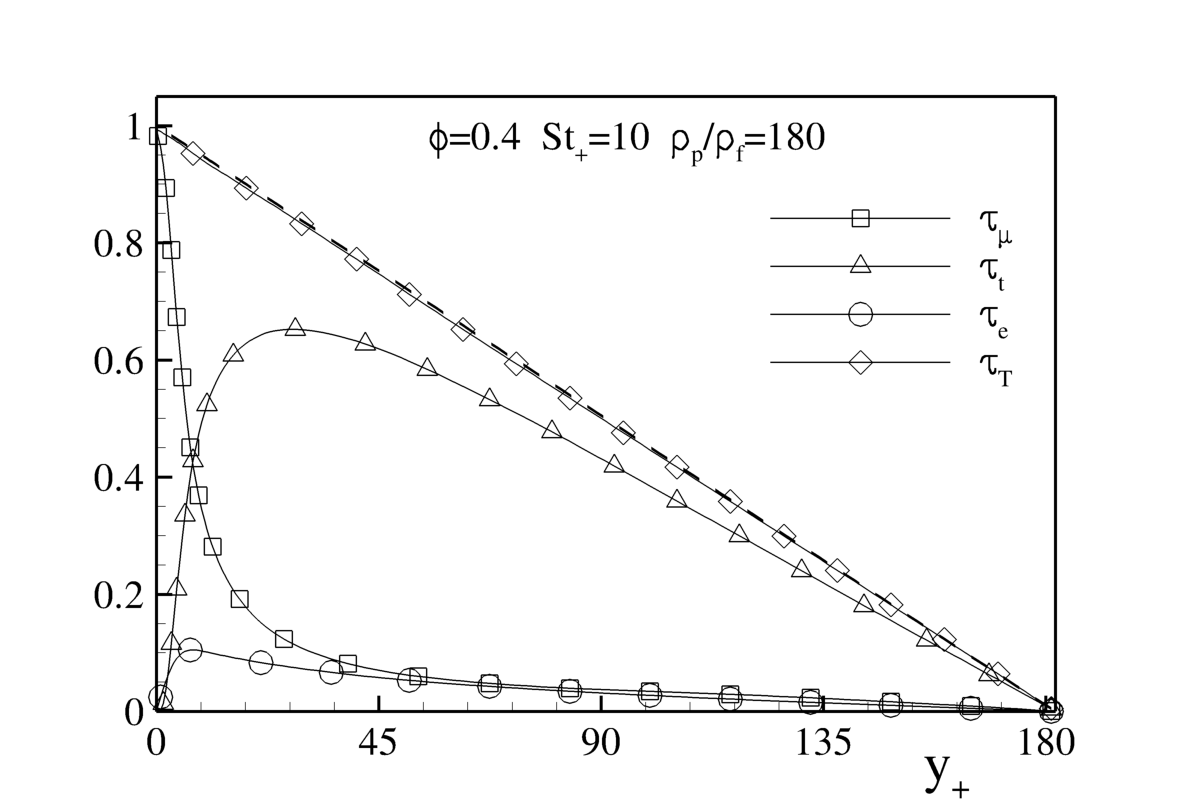}
{\scriptsize \put(-370,105){\bf (a)}}
{\scriptsize \put(-185,105){\bf (b)}}
\includegraphics[width=.48\textwidth]{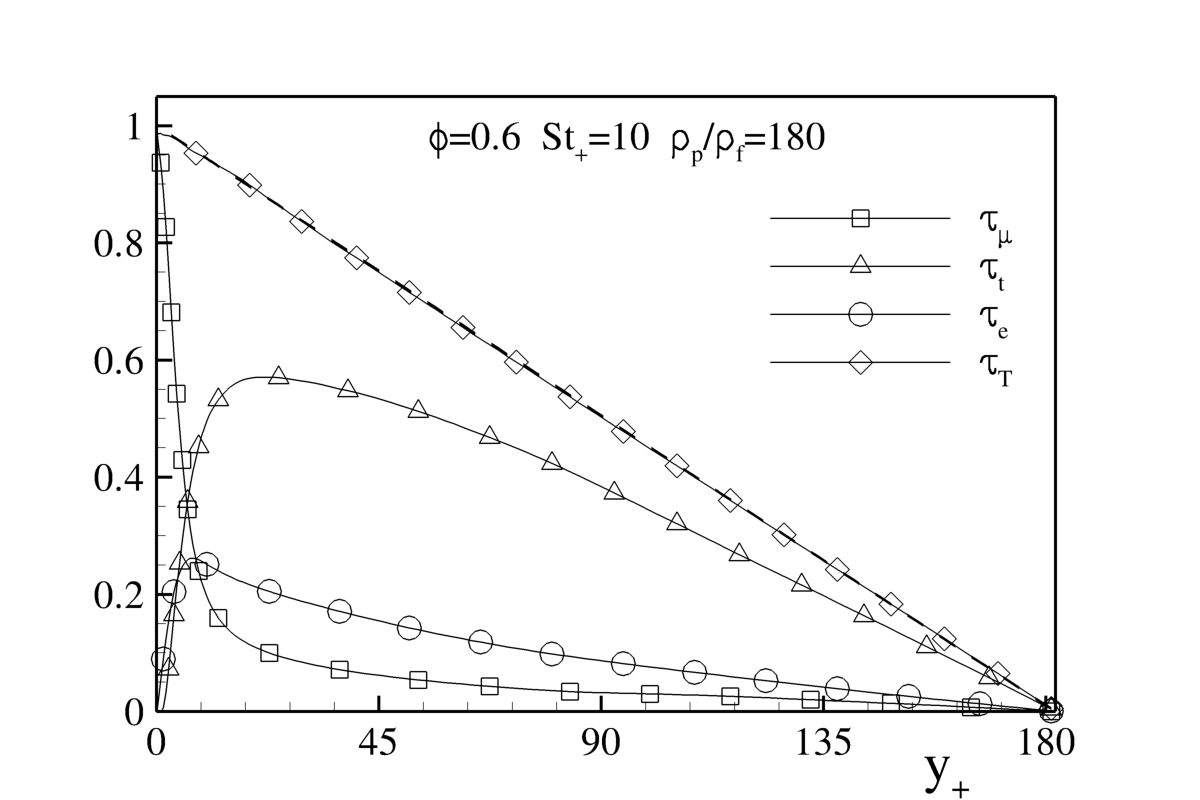}
\includegraphics[width=.48\textwidth]{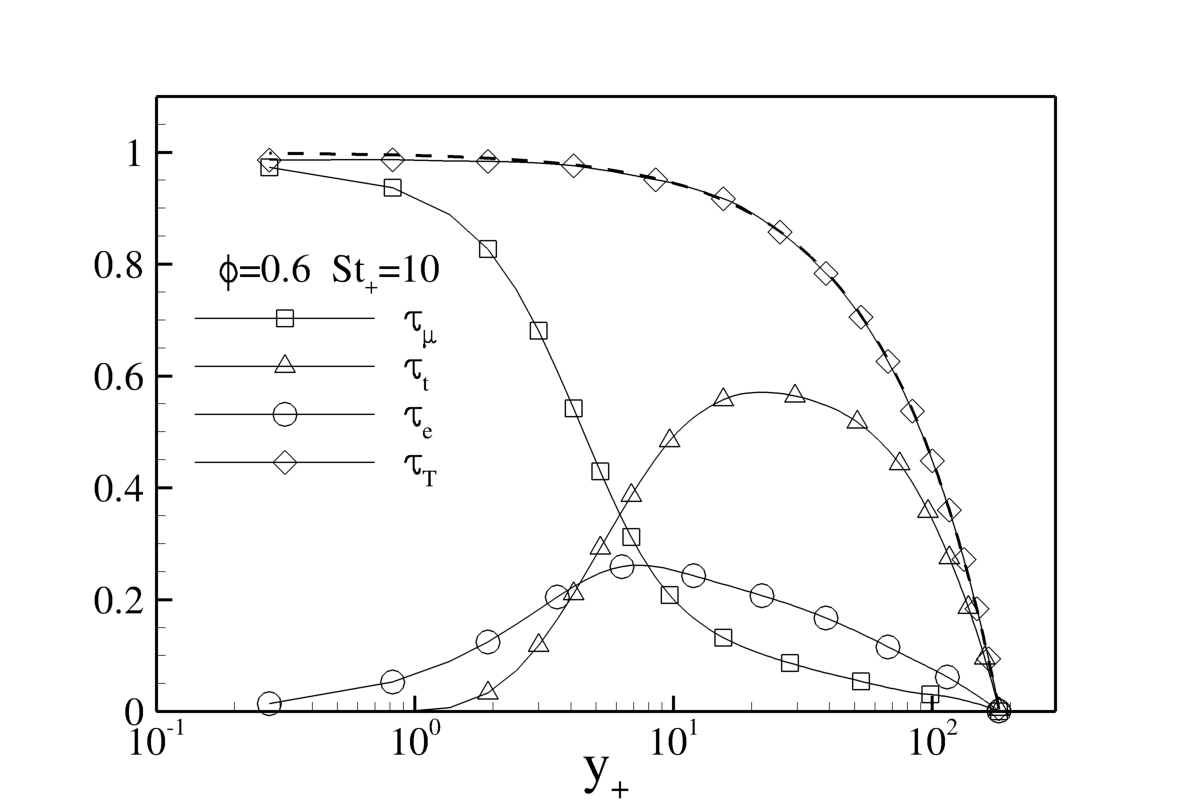}
{\scriptsize \put(-370,105){\bf (c)}}
{\scriptsize \put(-185,105){\bf (d)}}
}
\caption{Normalized mean stress balance eq.~\eqref{eqn:stress_balance} against  
wall-normal distance $y_+=(R-r)/y_*$. Viscous stress $\tau_\mu$
($\square$), turbulent stress $\tau_t$ ($\triangle$), extra stress
$\tau_e$ ($\bigcirc$), total stress $\tau_T=\tau_\mu+\tau_t+\tau_e$ 
($\meddiamond$) and $\displaystyle \frac{dp}{dz}\Big|_0 r$ (dashed line),
(see text for definitions). All stresses are normalised with the wall shear stress 
$\tau_w$. Panel a): uncoupled case; Panel b): case $\phi=0.4$ and 
$St_+=10$;  Panel c) case $\phi=0.6$ and $St_+=10$; d) same as panel c) but 
in semi-logarithmic scale.
\label{fig:stress_cfr}
}
\end{figure}
\begin{figure}
\centering{
\includegraphics[width=.48\textwidth]{./FIGS_FINAL/rev_stress_phi=04_St=10.png}
\includegraphics[width=.48\textwidth]{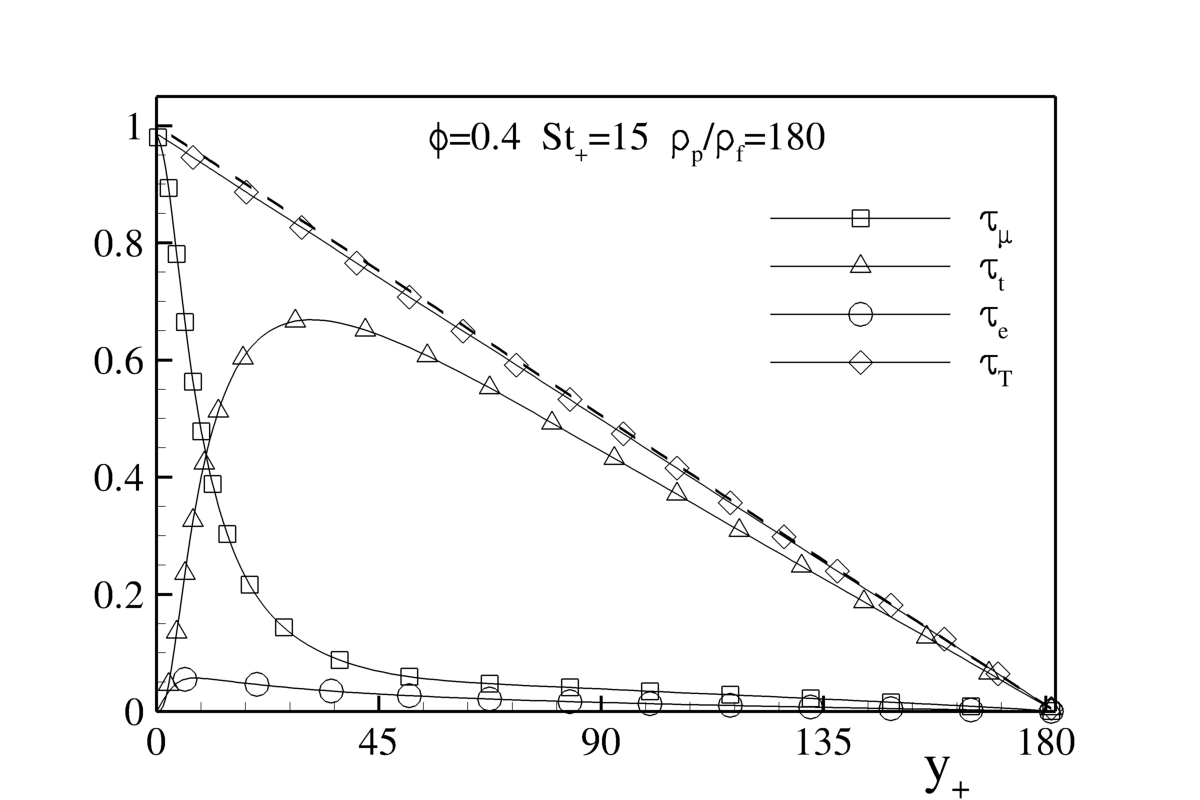}
{\scriptsize \put(-370,105){\bf (a)}}
{\scriptsize \put(-185,105){\bf (b)}}
\includegraphics[width=.48\textwidth]{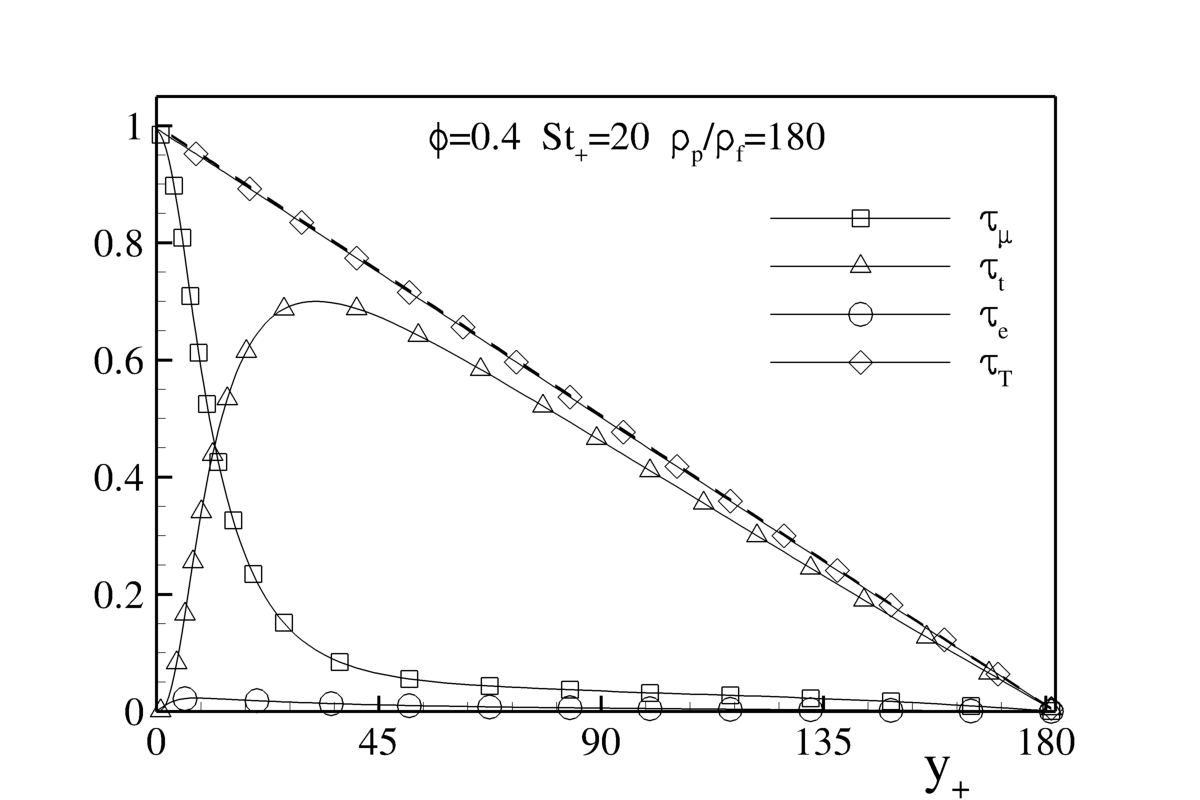}
\includegraphics[width=.48\textwidth]{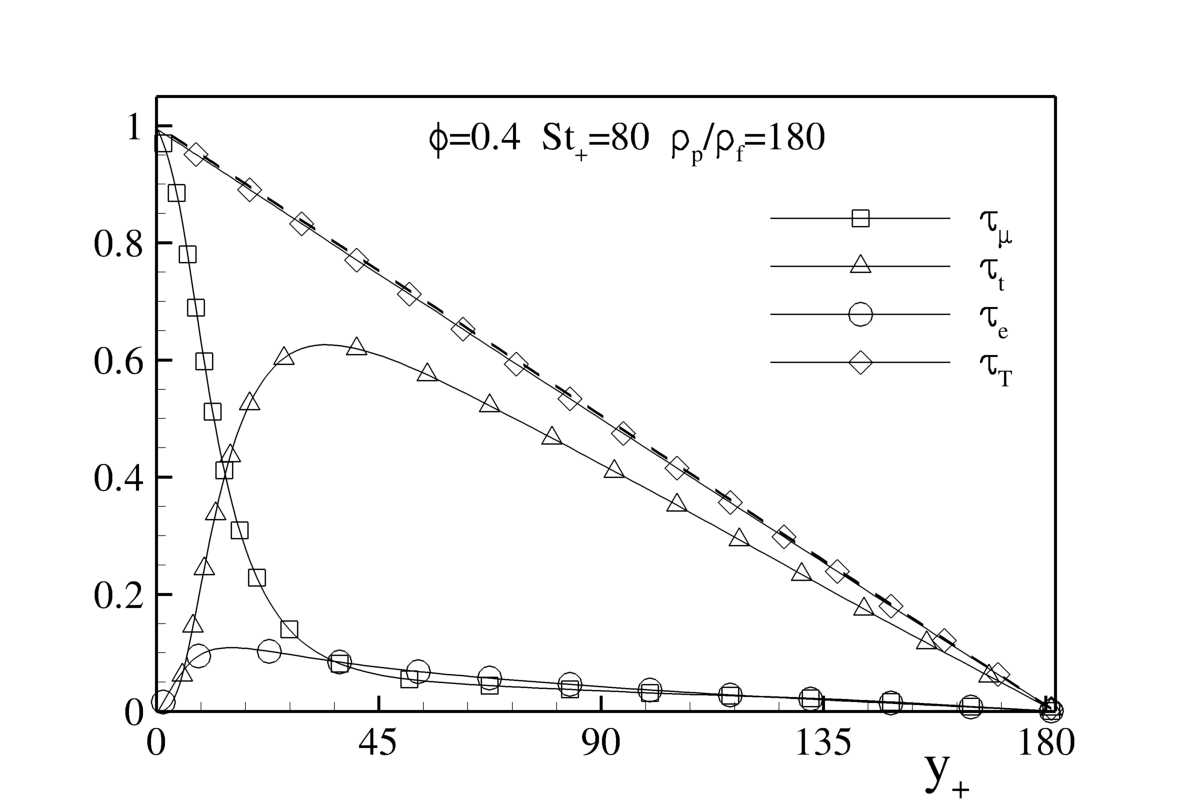}
{\scriptsize \put(-370,105){\bf (c)}}
{\scriptsize \put(-185,105){\bf (d)}}
}
\caption{Normalised mean stress balance eq.~\eqref{fig:stress_cfr} against the 
wall-normal distance $y_+=(R-r)/y_*$. Viscous stress $\tau_\mu$
($\square$), turbulent stress $\tau_t$ ($\triangle$), extra stress
$\tau_e$ ($\bigcirc$), total stress $\tau_T=\tau_\mu+\tau_t+\tau_e$ 
($\meddiamond$) and $\displaystyle \frac{dp}{dz}\Big|_0 r$ (dashed line),
(see text for definitions). All stresses are normalised with the wall shear stress 
$\tau_w$. Panel a) $\phi=0.4$, $St_+=10$; Panel b) $\phi=0.4$, $St_+=15$;
Panel c) $\phi=0.4$, $St_+=20$; Panel d) $\phi=0.4$, $St_+=80$.
\label{fig:stress_cfr_St}
}
\end{figure}
\begin{figure}
\centering{
\includegraphics[width=.48\textwidth]{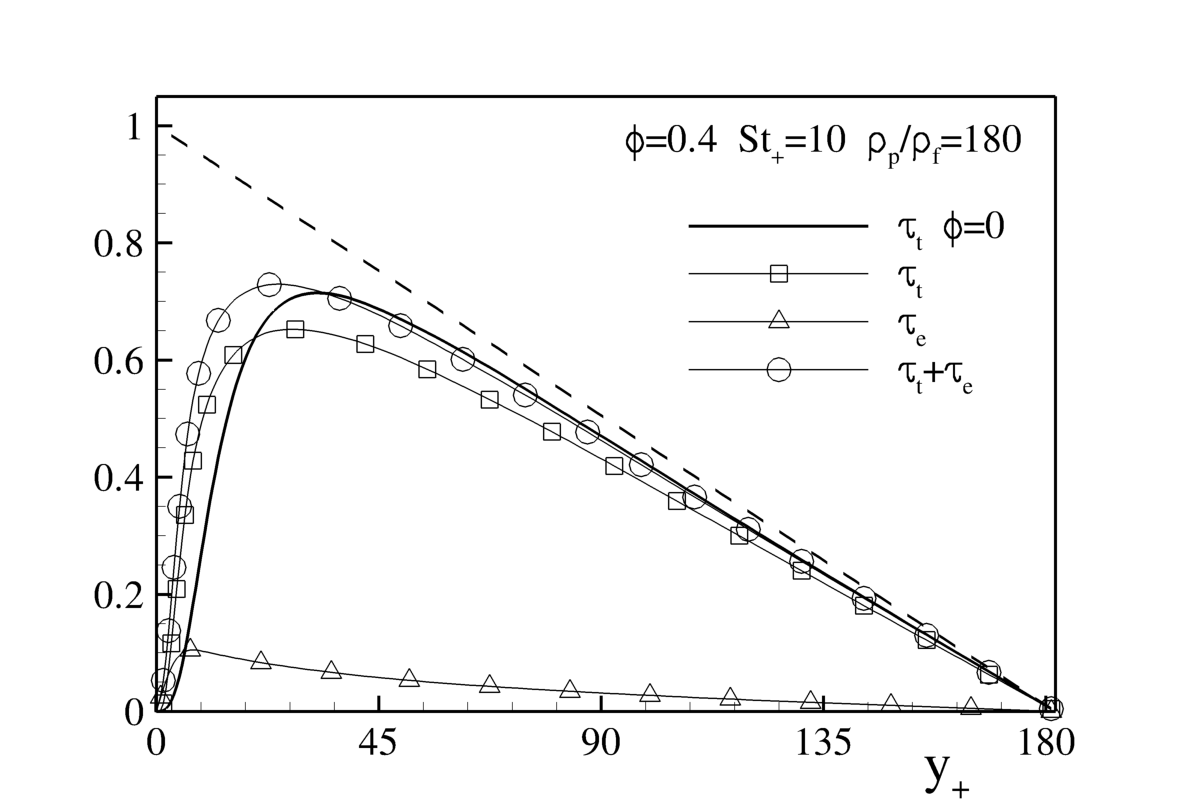}
\includegraphics[width=.48\textwidth]{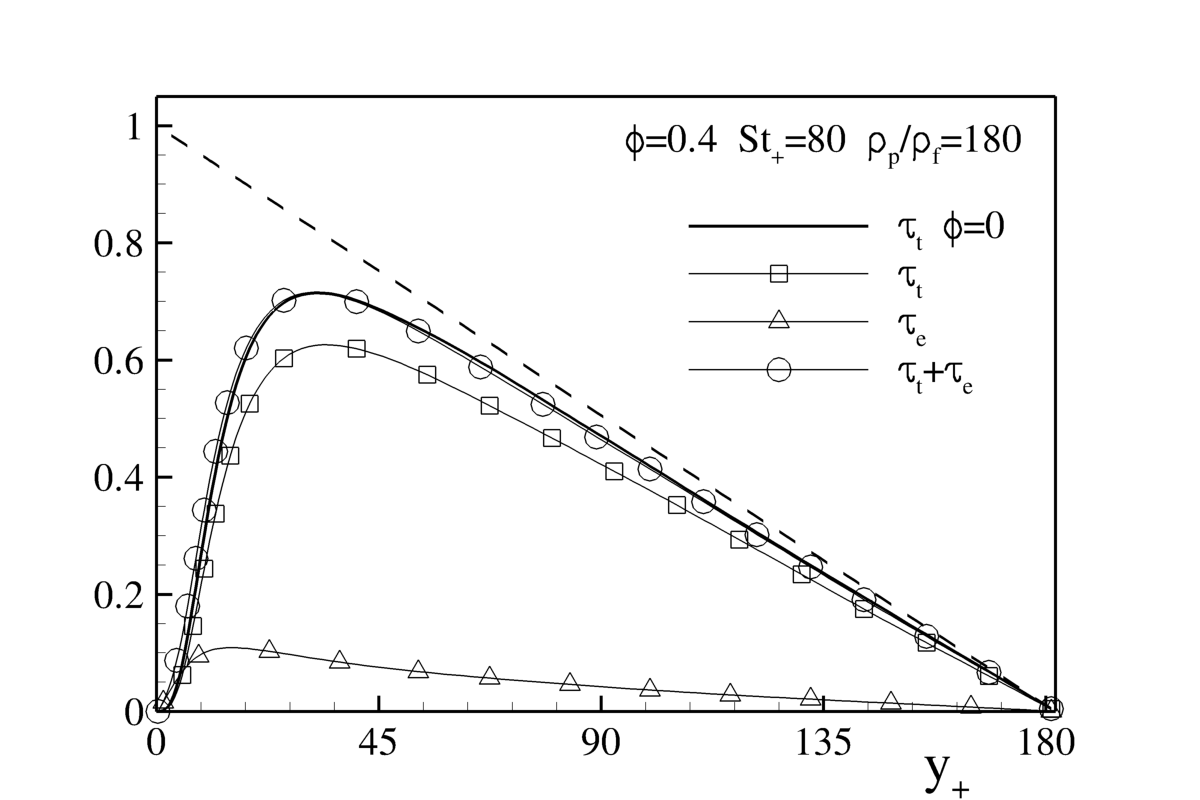}
{\scriptsize \put(-370,105){\bf (a)}}
{\scriptsize \put(-185,105){\bf (b)}}
}
\caption{Normalised mean stress balance eq.~\eqref{fig:stress_cfr} against the 
wall-normal distance $y_+=(R-r)/y_*$. Turbulent stress 
$\tau_t$ ($\square$), turbulent stress $\tau_t$ in the uncoupled case 
(solid line), extra stress $\tau_e$ ($\triangle$),
$\tau_t+\tau_e$ ($\bigcirc$), and $\displaystyle \frac{dp}{dz}\Big|_0 r$ 
(dashed line), see text for definitions. All stresses are normalised with the 
wall shear stress $\tau_w$. Panel a) $\phi=0.4$, $St_+=10$, $\rho_p/\rho_f=180$; 
Panel b) $\phi=0.4$, $St_+=80$, $\rho_p/\rho_f=180$; 
\label{fig:stress_cfr_oneway_St}}
\end{figure}

The particles in fully developed turbulent pipe flow modify the drag with respect to the 
uncoupled (Newtonian) case. This alteration non-trivially depends on mass loading, Stokes number and
particle-to-fluid density ratio.   Figure~\ref{fig:fric_coef} shows the friction coefficient
\begin{equation}
C_f = \frac{2\,\tau_w}{ \rho_f U^2} = \frac{\left. dp/dz\right|_0}{\rho_f U^2} \, ,
\label{eqn:fric_coef}
\end{equation}
where $U=Q/(\pi R^2)$ is the bulk velocity, $Q$ is the flow rate and $dp/dz|_0$ 
is the pressure gradient. In the figure $C_f$ is normalised with the unladen value, $C_{f0}$,  and plotted 
as a function of mass  loading (squares), Stokes number (circles) and particle-to-fluid 
density ratio (diamonds). Since the pressure drop is kept constant, an increase in friction 
coefficient corresponds to a decrease in mass flow rate.
 The figure shows that the drag increases at increasing mass loading
and decreases with Stokes number and density ratio. 
In the present range of parameters, the friction coefficient is always grater or at most equal to the 
uncoupled case value. 
\begin{figure}
\centering{
\includegraphics[width=.48\textwidth]{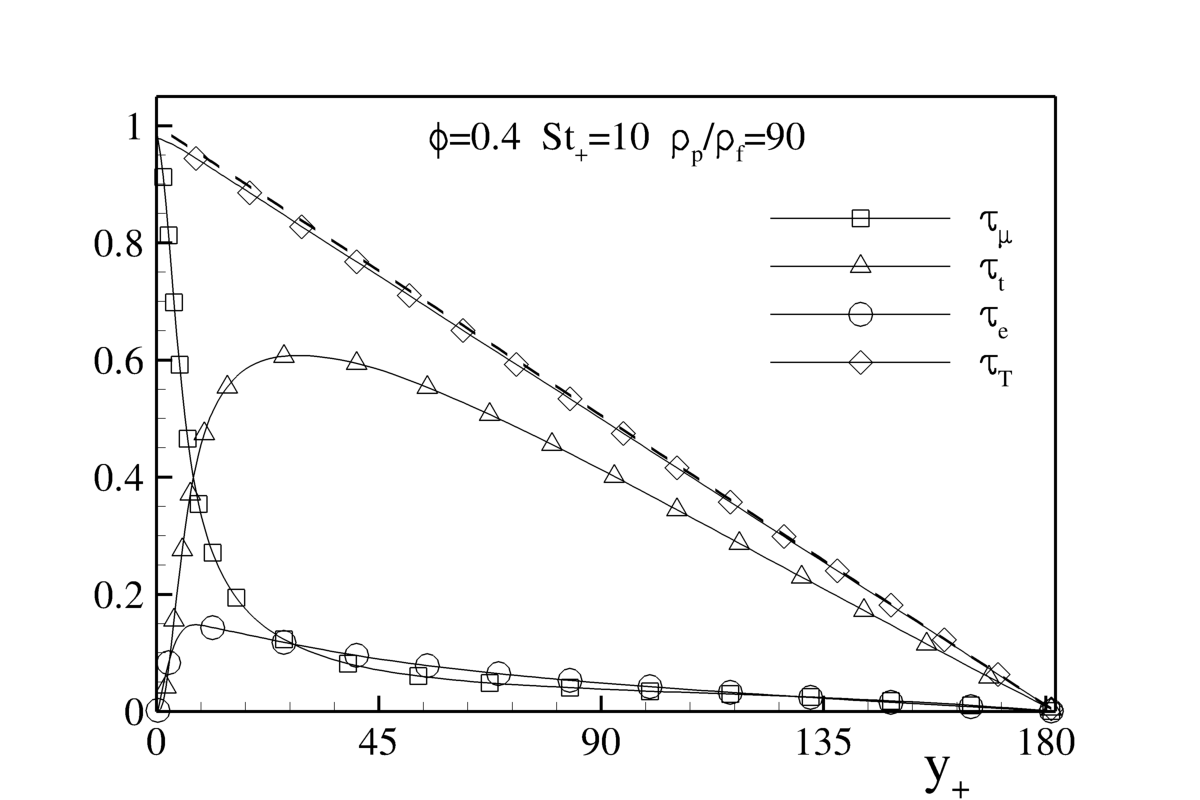}
\includegraphics[width=.48\textwidth]{./FIGS_FINAL/rev_stress_phi=04_St=10.png}
{\scriptsize \put(-370,105){\bf (a)}}
{\scriptsize \put(-185,105){\bf (b)}}
\includegraphics[width=.48\textwidth]{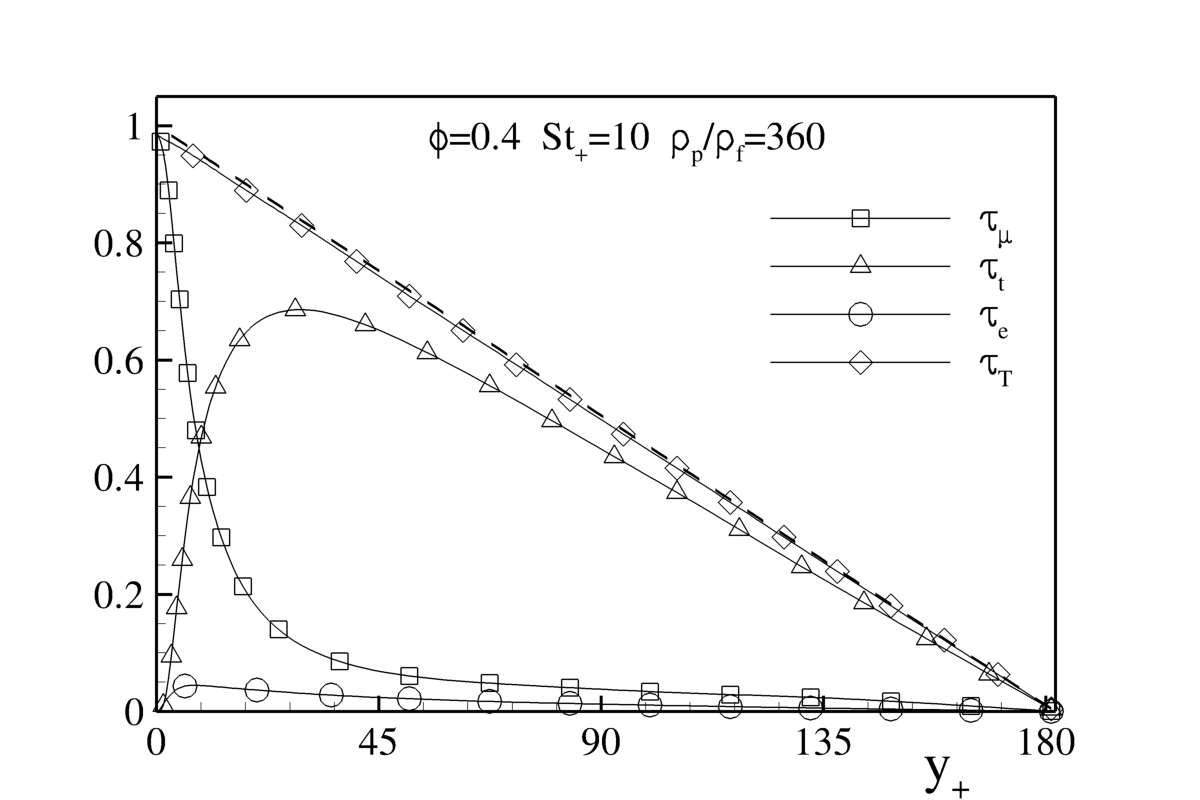}
\includegraphics[width=.48\textwidth]{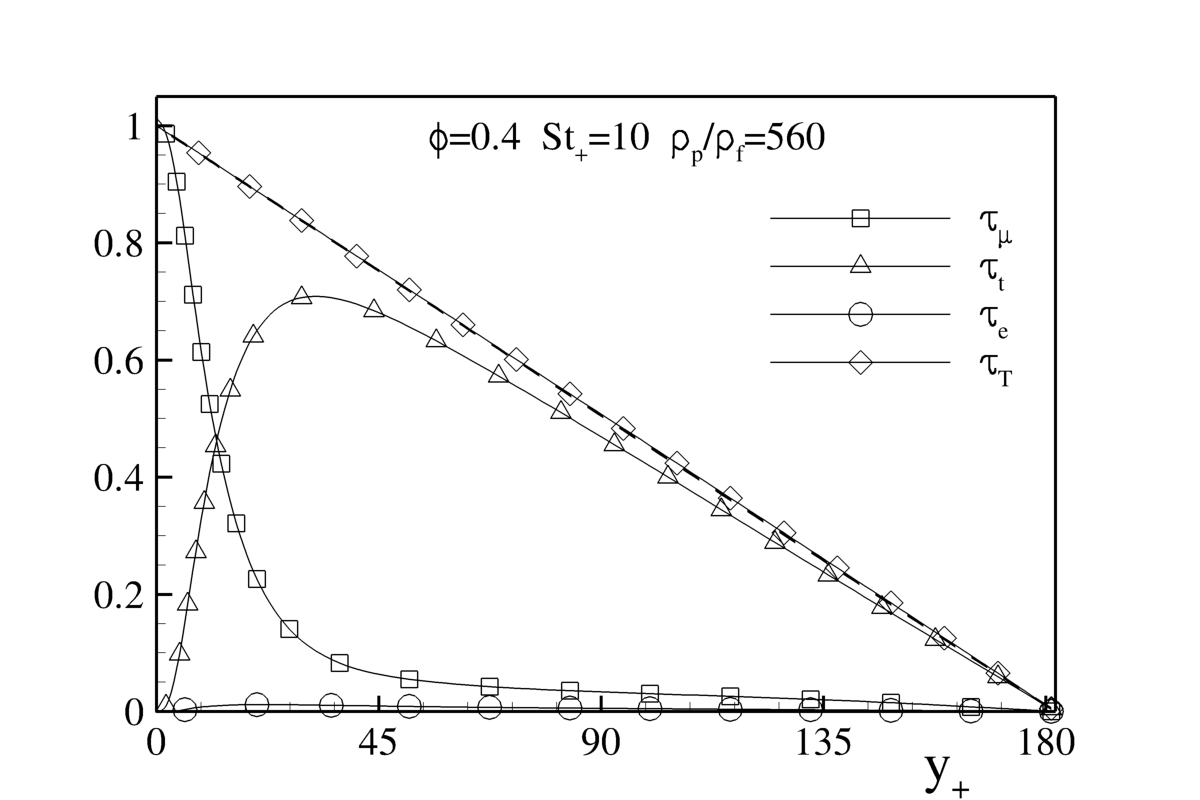}
{\scriptsize \put(-370,105){\bf (c)}}
{\scriptsize \put(-185,105){\bf (d)}}
}
\caption{Normalised mean stress balance 
eq.~\eqref{fig:stress_cfr} against the 
wall-normal distance $y_+=(R-r)/y_*$. Viscous stress $\tau_\mu$
($\square$), turbulent stress $\tau_t$ ($\triangle$), extra stress
$\tau_e$ ($\bigcirc$), total stress $\tau_T=\tau_\mu+\tau_t+\tau_e$ 
($\meddiamond$) and $\displaystyle \frac{dp}{dz}\Big|_0 r$ (dashed line),
(see text for definitions). All stresses are normalised with the wall shear 
stress $\tau_w$. In all panels $\phi=0.4$, $St_+=10$. 
Panel a) $\rho_p/\rho_f=90$; Panel b) $\rho_p/\rho_f=180$;
Panel c) $\rho_p/\rho_f=360$; Panel d) $\rho_p/\rho_f=560$.
\label{fig:stress_cfr_rho}
}
\end{figure}
\subsection{Mean momentum balance} \label{sec:mom_bal}
The drag modification is attributed to the alteration of the different contributions to the stress balance,
see e.g.~\cite{fukagata2002contribution}, 
\begin{equation}
\label{eqn:stress_balance}
\mu \frac{\partial U_z}{\partial r} -\rho_f\langle u'_r \, u'_z\rangle
+\frac{1}{r}\int_0^r \eta F_z \, d\eta =\frac{1}2{} \frac{dp}{dz}\Big|_0 r \, ,
\end{equation}
where $U_z=\langle u_z\rangle$ is the mean axial velocity,
$- \rho_f \langle u'_r \, u'_z\rangle$ is the turbulent Reynolds shear stress and 
$F_z=\langle f_z\rangle$ is the mean axial backreaction,
with angular brackets denoting ensemble average and primed variables representing fluctuations.
\begin{figure}
\centering{
\includegraphics[width=.6\textwidth]{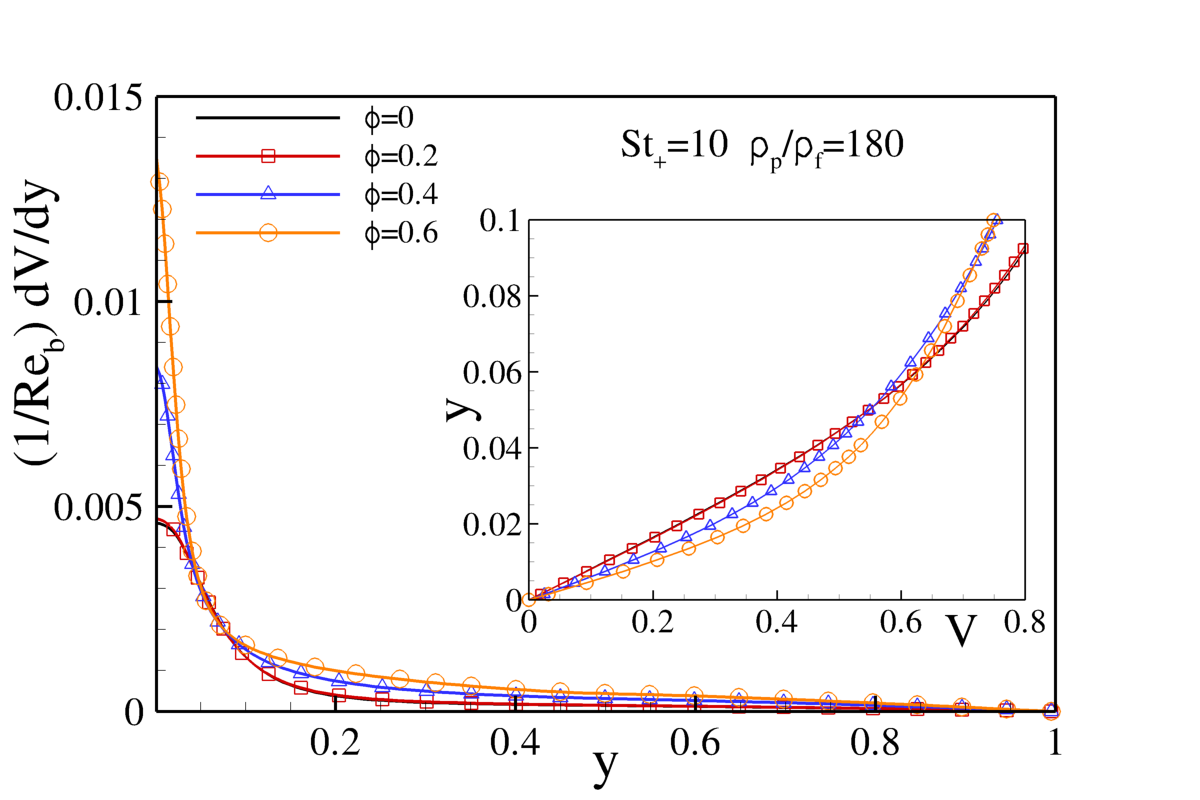}
{\scriptsize \put(-245,135){\bf (a)}}
\includegraphics[width=.6\textwidth]{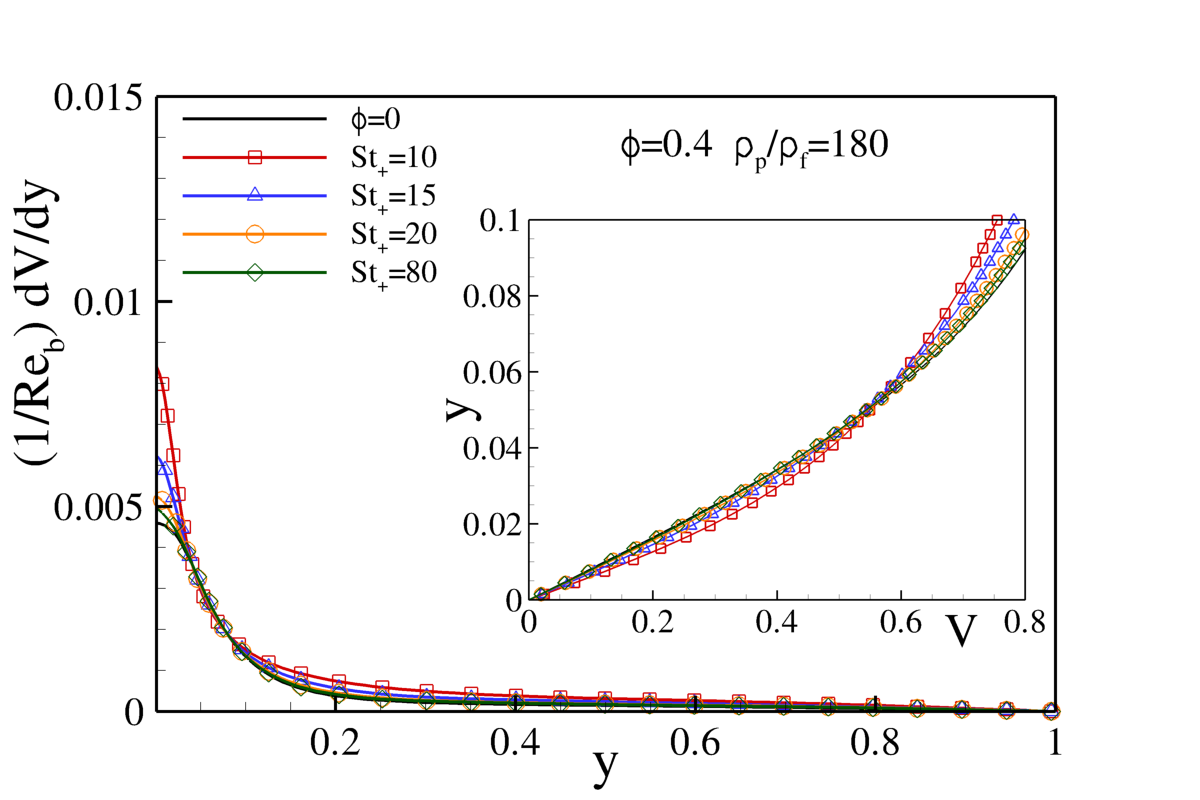}
{\scriptsize \put(-245,135){\bf (b)}}
\includegraphics[width=.6\textwidth]{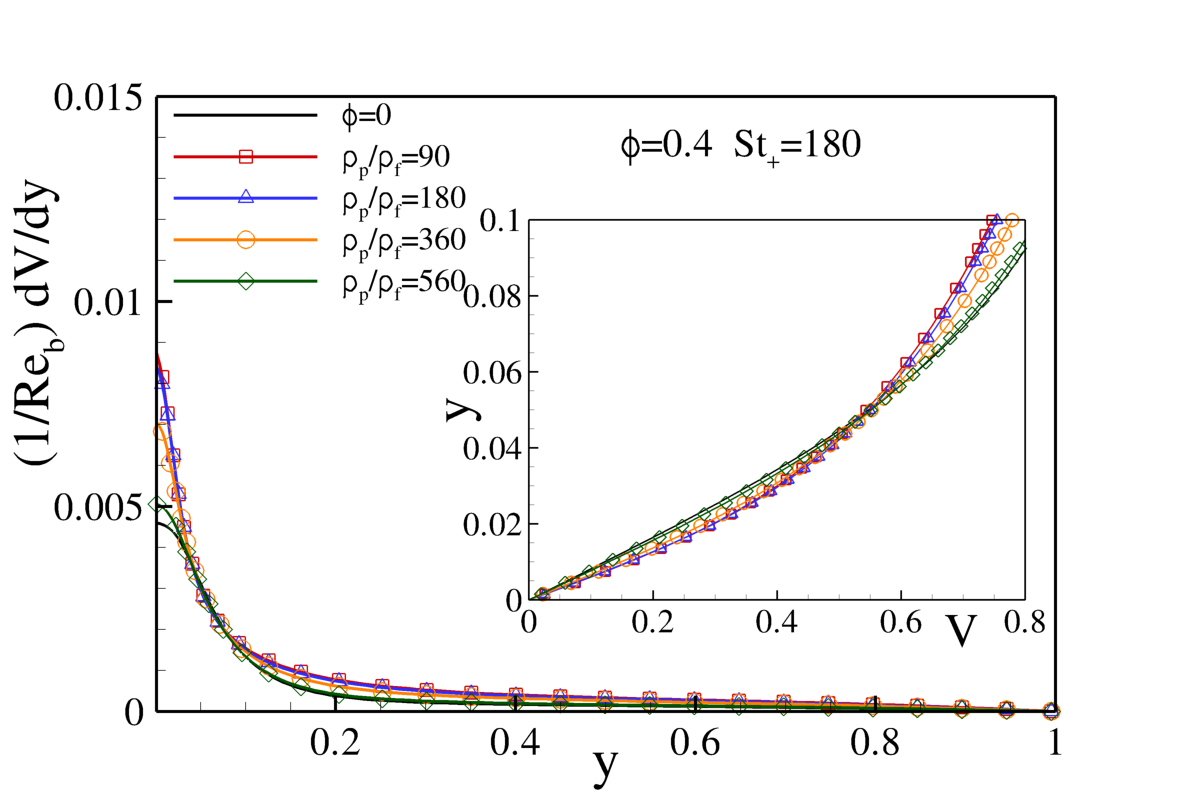}
{\scriptsize \put(-245,135){\bf (c)}}
}
\caption{Normalised mean viscous stress profiles against the 
wall-normal distance $y=(R-r)/R$. 
The inset report the mean velocity 
profile normalised with the bulk velocity, $V=U_z/U_b$, against the wall-normal distance close to the
wall.
\label{fig:viscous_stress}
}
\end{figure}
{In absence of particles,  the total shear stress, which is the sum of viscous stress,
$\tau_\mu = \mu \partial U_z / \partial r$, and of turbulent Reynolds shear stress, 
$\tau_t=-\rho_f \langle u'_r \, u'_z\rangle$, is a linear function of the radial coordinate.
This result can be derived by integrating once the mean axial momentum balance, 
see the classical textbook by \cite{pope2001turbulent}. Following the same procedure, 
in presence of a disperse phase, the particle-induced extra stress, 
$\tau_e=\frac{1}{r}\int_0^r \eta F_z \, d\eta$, arises
but the sum of the three stresses is still a linear function of the radial coordinate
as imposed by the global axial momentum balance.}
This should not be taken for granted in a DNS, unless some care is devoted to reach the statistically
steady state and acquire a well converged statistics. 
The critical cases correspond to large mass loading (large number of particles) and
large Stokes number and/or small density ratio (small number of particles). The former due to the large computational cost, the latter due to the long runs required to have converged statistics.

Figure~\ref{fig:stress_cfr} shows the balance of eq.~\eqref{eqn:stress_balance} for $\phi = 0$ (uncoupled), $\phi = 0.4$ and $\phi = 0.6$ in panels a), b) and c), respectively. 
The turbulent stress is attenuated almost everywhere and its peak shifts towards the wall. Indeed,
an extra stress arises that can be interpreted as an additional momentum flux towards the wall that modifies the turbulence dynamics and is at the  
origin of the drag increase. Its effect intensifies with increasing mass loading.

Figure~\ref{fig:stress_cfr_St} shows how the extra stress and the Reynolds shear stress 
profiles shift away from the wall at increasing Stokes number. 
As a consequence, the drag at $St_+=10$ is significantly higher than the 
drag at $St_+=80$ even though the extra stress has comparable values. 
Figure~\ref{fig:stress_cfr_oneway_St} goes deeper into the comparison with the uncoupled case, 
by considering $\tau_t+\tau_e$ at $St_+=10$ and $St_+=80$ and the turbulent Reynolds shear stress.
At $St_+=10$, the profile of $\tau_t+\tau_e$ peaks much closer to the wall and is more intense than the 
uncoupled turbulent Reynolds stress profile. On the other hand, the distribution of $\tau_t+\tau_e$ 
closely reproduces the uncoupled Reynolds stress at $St_+=80$. 
This combination further explains the difference in the drag between $St_+=10$ and $St_+=80$.

To complete the discussion about the friction coefficients, the 
effect of the particle-to-fluid density ratio on the stress contributions 
is shown in figure~\ref{fig:stress_cfr_rho}. The drag modification occurs since i) the extra stress 
decreases with increasing density ratio, becoming negligible at $\rho_p/\rho_f=560$ (the behaviour is 
similar at $\rho_p/\rho_f=900$ and is not shown), and ii) the 
turbulent Reynolds stress peak increases and departs from the wall region.

\begin{figure}
\centering{
\includegraphics[width=.48\textwidth]{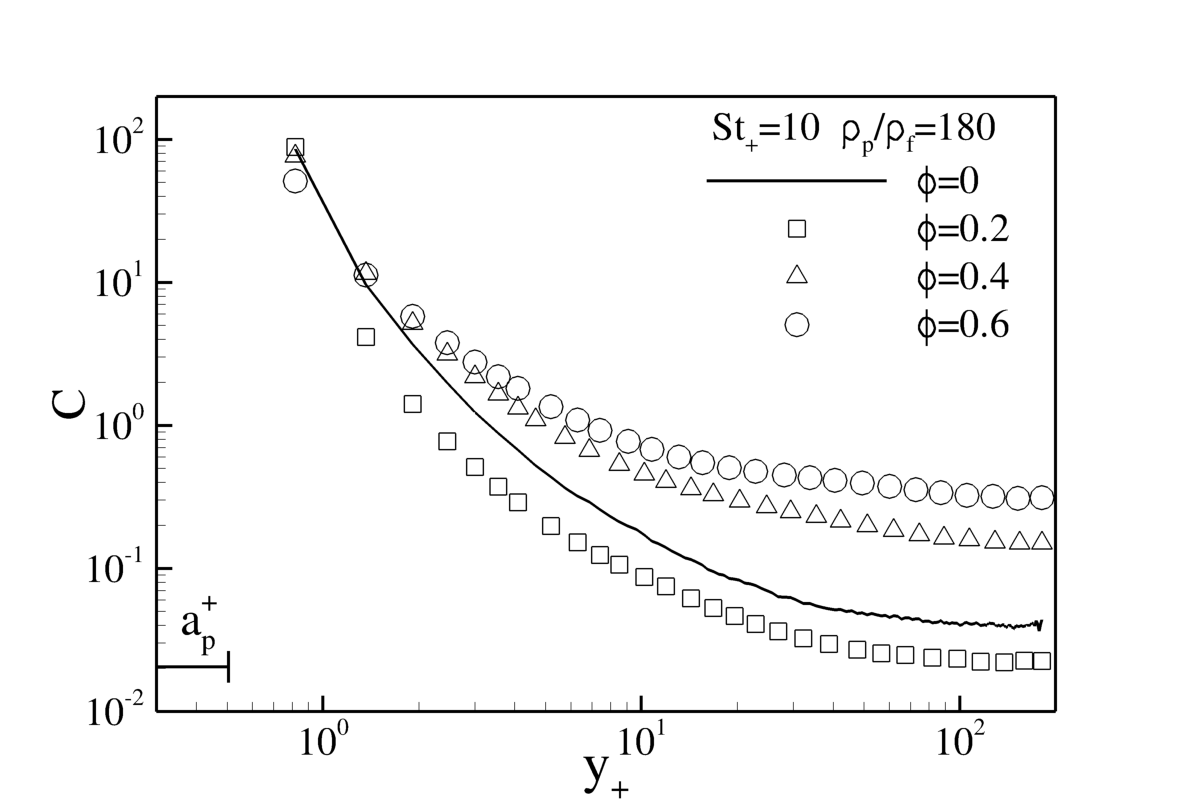}
\includegraphics[width=.48\textwidth]{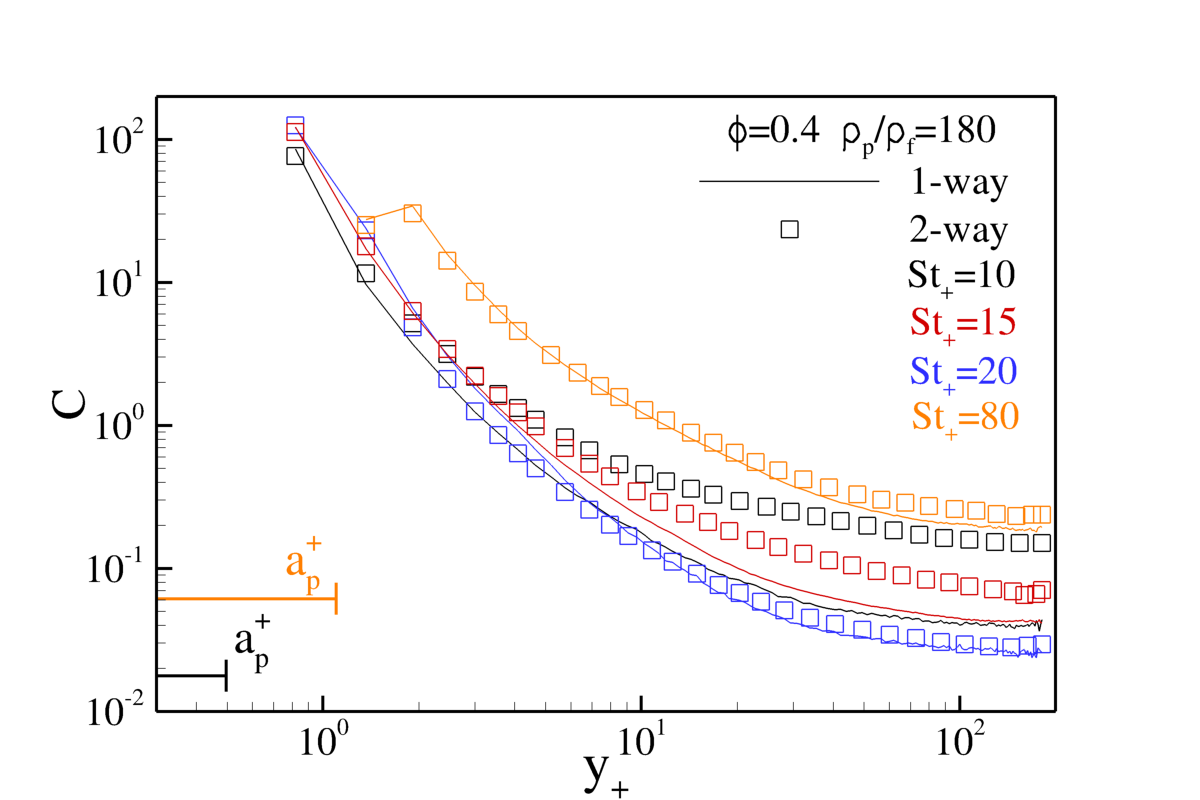}
{\scriptsize \put(-370,105){\bf (a)}}
{\scriptsize \put(-185,105){\bf (b)}}
\includegraphics[width=.48\textwidth]{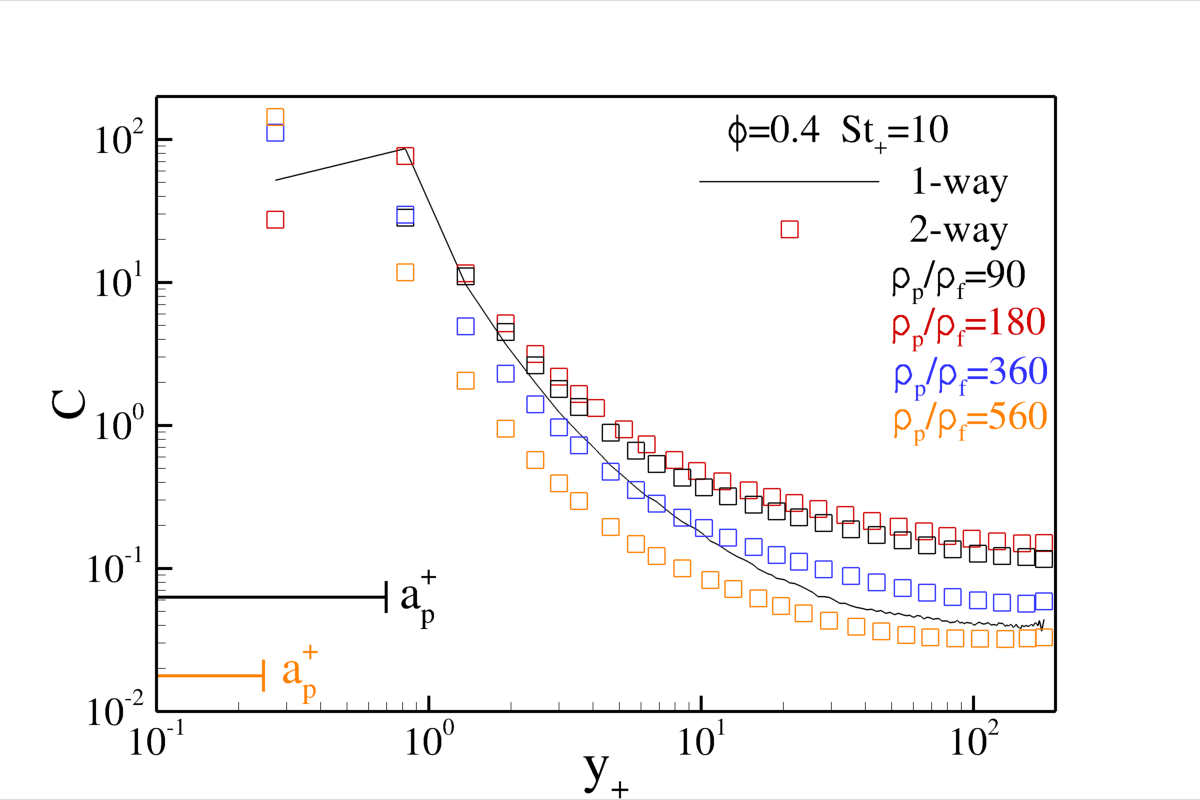}
{\scriptsize \put(-185,105){\bf (c)}}
}
\caption{Mean particle concentration $C$ against the wall 
normal distance $y=(R-r)/y_*$.
The lines refer the particle concentration in the uncoupled case whilst the 
symbols show the concentration in the two-way regime. The nominal 
particle radius is indicated with a horizontal line with the same color code of
the data in the plots.
Panel a): data at $St_+=10$ and $\rho_p/\rho_f=180$ and different mass load, 
$\phi=0.2$ ($\square$), $\phi=0.4$ ($\triangle$) and $\phi=0.6$ ($\bigcirc$). 
Panel b): effect of Stokes number, $St_+=10$ (black), $St_+=15$ (red), $St_+=20$ 
(blue) and $St_+=80$ (orange). Panel c): effect of the density ratio 
$\rho_p/\rho_f=90$  (black), $\rho_p/\rho_f=180$ (red), $\rho_p/\rho_f=360$ (blue),
$\rho_p/\rho_f=560$ (orange).
\label{fig:mean_concentration}
}
\end{figure}

The particles' feedback produces two concurrent effects: the depletion of the turbulent Reynolds shear 
stress and the presence of the particle extra stress. The latter produces an
increase of momentum flux towards the wall. The extra stress is then mainly balanced by the viscous 
stress since the turbulent Reynolds shear stress approaches zero. The modification of the viscous stress
results in a modification of the drag as shown in figure~\ref{fig:fric_coef}. 
To better highlight this behaviour, figure~\ref{fig:viscous_stress} reports the normalised mean viscous 
stress profiles as a function of the wall distance, with insets showing a closeup view of the normalised 
mean axial velocity profile. When the particle 
extra-stress provides a significant momentum flux towards the wall, the fluid velocity increases with 
respect to the uncoupled case. As a consequence, the viscous stress increases and the friction follows 
the same fate.  
{The main modification of the stress balance clearly occurs 
close to the wall. The extended ERPP method has been  designed to capture
the particle/fluid interaction close to a solid boundary accounting for the correct rate
of vorticity generation which, in turns, results in a physically consistent representation of 
the viscous shear stress and thus of the overall drag. This is a distinct characteristic
of the present approach which allows the prediction of the increase in drag.}
{\subsection{Mean particle concentration}}
\label{sec:mean_and_flucts_prtcls}
{
The particle mean distribution is presented in figure~\ref{fig:mean_concentration}
as a function of the wall-normal distance. The particle concentration is 
defined as $\displaystyle C(r) = (n_p/\Delta V_r)/(N_p/V_f)$,
where $n_{p}$ is the number of particles in a cylindrical shell of volume
$\Delta V_r=2\pi\, r\, \Delta r\, L_z$ placed at distance $r$ from the axis,
$N_p$ is the total number of particles in the fluid domain $V_f$.  
The normalisation of $C(r)$ is chosen such that $C=1$ when the 
particles are homogeneously distributed throughout the fluid domain.
In the one-way coupling regime, inertial particles tend to segregate in the near wall
region. The preferential accumulation, i.e. the turbophoresis, 
see~\cite{caporaloni1975transfer,reeks1983transport} 
and the review by~\cite{balachandar_rev}, is controlled by 
the Stokes number, see~\cite{marchioli2002mechanisms,sardina2012wall}
for the channel and the boundary layer respectively.
{
We have checked that the concentration profiles for the present simulations operated
in the one-way coupling regime, match the data reported by~\cite{picano2009spatial}
in a spatially developing pipe flow, and by \cite{sardina2011large} for a statistically steady 
pipe flow, by comparing the results when the flow has reached 
fully developed conditions (data not shown, Picano private communication).}
The question is whether the backreaction and the resulting 
turbulent modification is able to alter the particle accumulation across the pipe. \\
Figure~\ref{fig:mean_concentration} addresses the effect of (a) mass loading, (b)
Stokes number and (c) density ratio. At low mass loading ($\phi=0.2$) 
the particle concentration through the pipe decreases and particles segregate more at 
the wall.  The opposite occurs when the mass loading increases.
Concerning the Stokes number, the backreaction is effective in modifying 
the particle concentration with respect to the uncoupled case only for the populations 
at $St_+=10$ and $St_+=15$. Even at high Stokes number ($St_+=80$), the particles are still
unevenly distributed across the flow domain. In the previous section, negligible turbulence 
modification was seen for this Stokes number, therefore, accumulation of particles is not 
necessarily the only precursor for turbulence modification. Panel c) addresses the effect 
of the density ratio. The solid line, representing the concentration in 
the one-way coupling regime, is the same for all cases since it only depends  
on the Stokes number. The trend of particle concentration in the bulk of the flow 
is not monotonic, the highest being at $\rho_p/\rho_f=180$. 
The opposite behaviour is observed at the wall. 
Unlike the uncoupled case, the density ratio is a further crucial parameter in the two-way 
regime that influences the particle concentration.}

\section{Final remarks} \label{sec:final_remarks}

A proper methodology to account for the inter-phase 
momentum exchange between inertial particles and the carrier flow in presence
of wall has been developed. The approach extends the original ERPP method
to account for the additional physics introduced by the wall. 
The disturbance generated by small particles can still be evaluated in a closed 
form by considering the associated unsteady Stokes problem 
in the half-space where only the impermeability boundary condition
is enforced at the wall. When the disturbance is transferred to the background flow, 
the no-slip boundary condition is enforced by the Navier-Stokes solver of the carrier 
phase. From a physical point of view, this step corresponds to the generation and 
diffusion of the vorticity generated by the particles close to the wall. 
The approach has been carefully validated, showing how the impulse generated by the
particles is correctly transferred to the 
fluid impulse in the bulk and to the viscous drag force at the wall. These results 
highlight the need to consider a set of images for those particles that  
lie close to the wall in order to reproduce the correct physics of the inter-phase momentum
coupling.

The second part of the paper addresses the extended ERPP 
approach applied to direct numerical simulations of particle-laden fully developed 
turbulent pipe flow. The physical consistency of the inter-phase coupling method 
allows for a reliable analysis of the stress budget. In the near-wall region,
the ERPP approach has been proven to capture the
vorticity generated by the particles and the ensuing viscous shear stress. 
Results show a modification of the turbulent Reynolds shear
stress and the important role played by the extra stress 
produced by the particles close to the wall. The physical
interpretation corresponds to an augmented momentum flux towards the wall 
that ultimately increases the viscous shear stress and consequently the drag. 

{The approach applies to small particles, i.e. diameter comparable to  
the smallest hydrodynamical length-scale of the flow. The disturbance must be described by the 
unsteady Stokes equations, i.e. the particle Reynolds number is small. The suspension 
is considered diluted since inter-particles collisions and hydrodynamic interactions 
are neglected. No limitations are present on the density ratio, i.e. the approach can be used 
either for heavy particles or light bubbles. Clearly, in the latter case, added-mass and lift effects in 
the expression of the force on the bubble must be considered. }

\section*{Acknowledgements}
The research leading to these results has received funding from the European 
Research Council under the European Union's Seventh Framework Programme 
(FP7/ 2007-2013)/ERC Grant Agreement No. [339446].
\bibliographystyle{jfm}
\bibliography{erpp_wall}
\end{document}